\documentclass[screen]{acmart}
\usepackage{algorithm}
\usepackage{algpseudocode}
\usepackage{xspace}
\usepackage{enumitem}
\usepackage[capitalise, noabbrev, nameinlink]{cleveref}
\usepackage{todonotes}
\usepackage{makecell}
\usepackage{colortbl}
\usepackage{tabularray}
\usepackage[dvipsnames]{xcolor}
\usepackage{framed}
\usepackage[many]{tcolorbox} 
\usepackage{pdfpages}

\newcommand{\code}[1]{\hypertarget{code:#1}{\textsc{#1}}\label{code:#1}}
\newcommand{\coderef}[1]{\hyperlink{code:#1}{\textsc{#1}}}

\usetikzlibrary{calc}
\definecolor{labelColor}{RGB}{0,0,255}
\newcommand{\uilabel}[1]{\protect\tikz [font=\sffamily, baseline={($ (current bounding box.center) - (0,.3em) $)}] \fill[fill=labelColor] (0,0em) circle (0.6em) node[text=white] {#1};}

\definecolor{Silver}{rgb}{0.95,0.95,0.95}
\definecolor{main}{HTML}{5989cf}    
\definecolor{sub}{HTML}{cde4ff}     

\tcbset{
    sharp corners,
    colback = white,
    before skip = 0.2cm,    
    after skip = 0.5cm      
}                           

\newtcolorbox{bluebox}{
    enhanced, 
    boxrule = 0pt, 
    borderline = {0.75pt}{0pt}{main}, 
    borderline = {0.75pt}{2pt}{sub} 
}

\newcommand{\yes}{$\bullet$} 
\newcommand{\partialf}{$\circ$}
\newcommand{\rotheader}[1]{\rotatebox{90}{\parbox{5cm}{\raggedright #1}}}

\definecolor{quotemark}{gray}{0.75}
\makeatletter
\def\fquote{%
    \@ifnextchar[{\fquote@i}{\fquote@i[]}
           }%
\def\fquote@i[#1]{%
    \def\tempa{#1}%
    \@ifnextchar[{\fquote@ii}{\fquote@ii[]}
                 }%
\def\fquote@ii[#1]{%
    \def\tempb{#1}%
    \@ifnextchar[{\fquote@iii}{\fquote@iii[]}
                      }%
\def\fquote@iii[#1]{%
    \def\tempc{#1}%
    \noindent%
    \begin{list}{}{%
         \setlength{\leftmargin}{0.06\textwidth}%
         \setlength{\rightmargin}{0.05\textwidth}%
                  }%
         \item[]%
         \begin{picture}(0,0)%
         \put(-8,-5){\makebox(0,0){\scalebox{3}{\textcolor{quotemark}{``}}}}%
         \end{picture}%
         \begingroup\itshape}%
 \def\endfquote{%
 \unskip\hspace{0.5em}%
 \makebox[0pt][r]{%
 \hspace{-0.05\textwidth}%
 \begin{picture}(0,0)(0,0)%
 \put(0,0){\makebox(0,0){%
 \scalebox{3}{\color{quotemark}''}}}%
 \end{picture}}%
 \endgroup%
 \ifx\tempa\empty%
 \else%
    \ifx\tempc\empty%
       \hspace{10pt} \tempa\ifx\tempb\empty\else,\ \emph{\tempb}\fi%
   \else%
       \hspace{10pt} \tempa,\ \emph{\tempb},\ \tempc%
   \fi\fi%
 \end{list}%
}%

\AtBeginDocument{%
  }

\acmISBN{978-1-4503-XXXX-X/18/06}

\begin{document}

\title{Teaching Geometric Proof with Tech: Pitfalls and Possibilities}

\author{Hwei-Shin Harriman}
\orcid{0000-0002-3746-4808}
\affiliation{%
  \institution{Carnegie Mellon University}
  \city{Pittsburgh}
  \state{PA}
  \country{USA}
}
\email{harriman@cmu.edu}

\author{Wode Ni}
\orcid{0000-0002-5341-4958}
\affiliation{%
  \institution{Carnegie Mellon University}
  \city{Pittsburgh}
  \country{USA}
}
\email{nimo@cmu.edu}

\author{Yuchen Jin}
\affiliation{%
  \institution{University of Wisconsin-Madison}
  \city{Madison}
  \country{USA}}
\email{rainyjin2016@gmail.com}

\author{Dominik Moritz}
\orcid{0000-0002-3110-1053}
\affiliation{%
  \institution{Carnegie Mellon University}
  \city{Pittsburgh}
  \country{USA}
}
\email{domoritz@cmu.edu}

\author{Joshua Sunshine}
\orcid{0000-0002-9672-5297}
\affiliation{%
 \institution{Carnegie Mellon University}
 \city{Pittsburgh}
 \country{USA}}
 \email{sunshine@cs.cmu.edu}

\renewcommand{\shortauthors}{Harriman et al.}

\begin{abstract}

Geometric proof is a foundational yet challenging topic in mathematics, requiring students to integrate visual, logical, and notational skills. While technology has enhanced learning in other mathematical domains, its impact on geometric proof remains limited. To investigate this gap, we interviewed 18 geometry teachers to establish the technical requirements of educational proof tools. These requirements inform our review of 33 commercial and research tools. Our findings reveal a critical mismatch: while teachers value certain digital tools for initial planning and exploration activities, they revert to pen-and-paper for formal proof because it supports diagram annotation and provides space for multiple approaches to proof-solving. Annotating the diagram is a key component of the proof-solving workflow that existing tools do not support. We propose four technical and human-centered design guidelines for educational proof tools to meet teacher needs at scale: integrating diagram and proof, generating problems and feedback automatically, supporting multiple proof formats, and reducing accidental complexity in the user experience.

\end{abstract}





\maketitle


\section{Introduction}\label{sec:intro}

Proof is the heart of mathematics; it is the mechanism that transforms conjecture into established fact. Geometry traditionally serves as the primary entry point for formal deductive reasoning. Its inherent visual nature provides a tangible bridge between intuitive observation and abstract logic. Despite its importance, geometric proof remains one of the most difficult topics for students to master, as it requires the simultaneous integration of spatial, logical, and symbolic skills~\cite{stylianou2009teaching, usiskin_van_1982, senk_how_1985}. To address these instructional challenges, many educators look toward digital tools. In other areas of mathematics, such as arithmetic or algebra, educational technology has successfully improved learning outcomes by offering immediate feedback, interactivity, and personalization~\cite{li2010meta}. However, these benefits have not translated to geometric proof with the same degree of success. We hypothesize that this discrepancy exists because proof instruction requires a specialized approach---distinct even from other mathematical domains---and current tools are not designed to provide comprehensive support.

The success of any intervention in this space is ultimately determined by the teacher, who acts as the ``gatekeeper'' of the classroom experience. Even when using standardized textbooks, teachers pick and choose specific problems and activities that shape how students encounter the material~\cite{sears2014opportunities}. Because teachers influence how geometric proof is taught, tools should reflect their classroom-tested beliefs. Accomplishing this requires us to first identify what teachers value and then evaluate whether current tools are actually supporting their values. This motivation leads to our first two research questions: 

\begin{enumerate}[label=\textbf{RQ\arabic*}]
    \item\label{rq:1} What approaches do teachers believe most improve their students' skills in geometric proof?
    \item\label{rq:2} To what extent do the tools that geometry teachers use support a comprehensive proof-solving workflow?
\end{enumerate}

To answer these first two research questions, we conducted 18 semi-structured interviews with geometry teachers in the United States. Our participants bring varied perspectives with an average of 19 years of experience (min of 3, max of 35), teaching a wide range of student skill levels at public and private high schools. Our interviews expose a mismatch between the needs of the teacher and the tools they have used. In \cref{sec:results} and \cref{sec:results-rq2}, we find that teachers value existing tools for specific tasks such as exploration and hypothesis-formation. However, most of their proof-specific instruction still relies on traditional pen-and-paper activities. They specifically value pen-and-paper because it enables students to add marks and annotate the proof as they work. Additionally, the proof-specific tools our participants had used suffered from various shortcomings. Among other issues, we found that the tools did not support diagram annotation, provided limited feedback to students, and oversimplified the proving process. The interviews reveal several topics of dissatisfaction, but it is possible that a tool our participants have never used already addresses these issues. This motivates our third research question:

\begin{enumerate}[label=\textbf{RQ3}]
    \item\label{rq:3} How well does the landscape of existing tools address the needs of teachers?
\end{enumerate}

\noindent To bridge the gap between teacher needs and available technology, we reviewed 33 tools (\cref{sec:tools}). We included both commercial tools mentioned by our interview participants and research tools found through literature review. The tools include intelligent tutoring systems, dynamic geometry environments, large language models, and automatic theorem provers. By mapping features to teacher needs, we found that while some tools excel in isolation, none offer comprehensive support. The largest design oversight concerns the role of the geometric diagram. Teachers expect their students to refer to and annotate the diagram at every step of the proof. Yet, our review shows that proof-specific tools consistently fail to provide the necessary digital affordances to support this visual reasoning.

Based on these findings, we suggest four technical and human-centered improvements that could close the gap between the features of the tools and the needs of the teacher (\cref{sec:discussion}). These improvements target the integration between the diagram and the proof and the creation of high-quality proof content. Reducing the manual labor required to create proof problems also opens up opportunities to explore new types of activities. Lastly, improving the user experience of both existing and future proof education tools would benefit teachers and students alike.


\section{An Example Geometric Proof}
\label{sec:example}
\begin{figure*}
    \centering
    \includegraphics[width=1\textwidth]{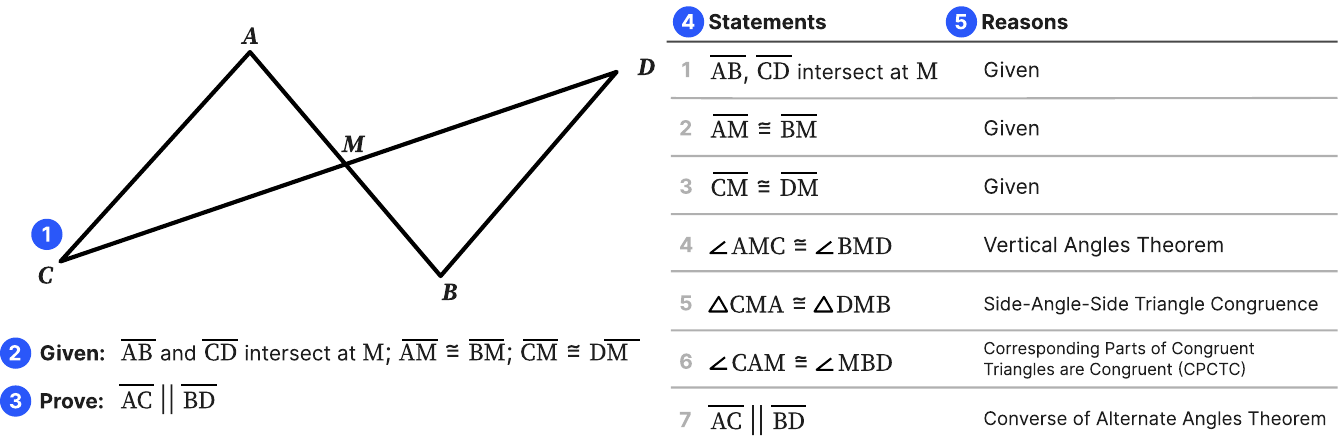}
    \caption{\textbf{Left:} Typical information provided to specify a proof problem, including a diagram construction \uilabel{1}, givens \uilabel{2} (statements assumed to be true for this proof), and the goal to prove \uilabel{3}. \textbf{Right:} Completed proof in two-column format, consisting of a series of statements \uilabel{4} justified by reasons \uilabel{5}. Reasons must either be ``given'' by the problem, or be other proven mathematical facts.}
    \Description{A two-part figure. On the left is a diagram illustrating the typical information provided for a proof problem, including a geometric construction, given statements, and the goal to be proven. On the right is a completed proof presented in two-column format, with statements in one column and their corresponding justifications in the other. The justifications are either ``given" from the problem or derived from mathematical facts such as theorems or axioms.}
    \label{fig:ex-proof}
\end{figure*}

This section delineates the core components of a geometric proof to provide a consistent reference point for the discussion that follows. \cref{fig:ex-proof} provides a concrete example of geometric proof structure to ground our terminology. The premises, shown on the left, include three pieces: the \textbf{construction} \uilabel{1} which illustrates the objects in the proof, \textbf{given} information \uilabel{2}, and the goal to \textbf{prove} \uilabel{3}. On the right, the two-column format lists a series of \textbf{statements} \uilabel{4} and \textbf{reasons} \uilabel{5}. Each statement must be justified by a reason, and each reason must be either given or a known geometric rule (e.g., theorem, definition). 

This proof has a logical flow commonly found in high school-level geometric proofs, where triangle congruence is proven to demonstrate that another property holds. Congruence means that objects are the same shape and size, and is denoted by the ``$\cong$'' symbol seen in statements 2 through 6. In this case, it first proves the two triangles are congruent by Side-Angle-Side triangle congruence (SAS), then applies the ``Corresponding Parts of Congruent Triangles are Congruent'' rule (CPCTC) to prove that $\angle CAM$ and $\angle MBD$ are congruent. These angles are then used as alternate interior angles to prove that the segments $\overline{AC}$ and $\overline{BD}$ are parallel.

\section{Background and Related Work}\label{sec:related}
This section outlines the topics necessary to engage with our interview results and tool review. First, we provide an overview of the challenges of learning proof. Next, we describe the role of the teacher as both curator and mediator who adapts textbook and technological resources to fit the needs of their curriculum. Finally, we introduce what is known about how existing tools impact student learning of proof. A deeper dive into existing technology is discussed in \cref{sec:tools}.  

\subsection{Challenges of learning proof}

Students consistently struggle to master the complexities of geometric proof~\cite{usiskin_van_1982,senk_how_1985}. This difficulty was quantified as early as 1985, when the National Council of Teachers of Mathematics reported that only 30\% of geometry students could produce acceptable solutions to proof-based problems~\cite{senk1985well}. The root of this difficulty lies in the specific cognitive habits students bring to the classroom. Conditioned by the procedural nature of prior mathematical instruction, students struggle to pivot to the rigorous deductive reasoning required for proof. This transition represents a cognitive shift that remains exceptionally challenging for high school learners~\cite{stylianou2009teaching}. This new way of thinking is daunting even for more advanced students: undergraduates often struggle to identify an appropriate starting point for a proof~\cite{moore1994making} or fail to understand which definitions and theorems they can invoke~\cite{weber2001student}.

Geometric proof is unique in that it integrates a diagram with formal deduction. This can help students visually track information, yet the diagram's role is deceptively complex. Certain low-level properties in a proof---such as between-ness or intersection---can be drawn directly from the diagram, whereas higher-level properties---such as congruence or perpendicularity---can only be concluded using a deductive argument~\cite{laborde2005hidden}. This distinction tempts students into placing blind faith in the diagram. As a result, many students incorrectly assume it is drawn to scale and derive conclusions based on relationships that visually appear to be true~\cite{Soucy_McCrone_Martin_2004a,cirillo2019addressing}. Furthermore, students often believe that a proof only applies to the specific diagram paired with it~\cite{Chazan_1993}. This limited perspective prevents students from grasping the full generality of proof.

The format of the proof itself is also known to contribute to misunderstandings. The two-column proof, as shown in \cref{fig:ex-proof}, is the most common format for geometric proof. However, its linear progression simplifies the underlying graph-shaped logical structure of a proof~\cite{ball_teaching_2003}. This format tempts students to view proof-solving as a sequential, procedural exercise. In reality, the same proof can be approached through several different logical paths.

\subsection{Teacher as a mediator of curriculum and technology}
The teacher serves as the primary guide for students navigating the difficulties of geometric proof. This role requires specialized expertise, encompassing everything from constructing valid arguments to evaluating the subtle logic of student solutions~\cite{lesseig2016investigating}. The teacher's pedagogical choices also dictate the students' opportunities to hone their reasoning skills~\cite{martin2005interplay}. When teachers pose open-ended tasks and provide iterative feedback on student conjectures, they foster an environment where students can actively engage in the deductive process.

This pedagogical ideal is often constrained by the materials available in the classroom. The interaction between these materials and the teacher is complex, as textbooks frequently emphasize memorization and low-level cognitive tasks~\cite{sears_examination_nodate}. Such procedural focus stands at odds with the open-ended reasoning required for proof, forcing teachers to adapt materials to meet their goals~\cite{sears2014opportunities}. This discretion extends into technology-rich environments. Even when using digital tools, the teacher decides when to introduce specific software, how to frame each task, and what methods students should use to achieve learning objectives~\cite{drijvers2010teacher}.

\subsection{Pedagogical reality of digital proof tools}
Digital tools have emerged as a central theme in geometry education research, with particular emphasis placed on \emph{Dynamic Geometry Environments} (DGE), such as \textsc{GeoGebra}~\cite{geogebra}, \textsc{Geometer's Sketchpad}~\cite{gsp_5}, or \textsc{Desmos}~\cite{desmos}. These environments allow students to construct, manipulate, and measure geometric figures. Most research about geometric tools focuses on the use and impact of these DGEs~\cite{sinclair2012technology, weigand2025geometry}. DGEs are useful for exploring or inducing the properties of a figure and forming hypotheses about whether such properties are provable~\cite{laborde2002integration}. These tools also improve student understanding of geometric shapes and their ability to perform mental manipulations of those shapes. However, they are not designed to support the student through formal deductive reasoning. This gap is illustrated by \citet{komatsu2020interplay}, who found that students, when provided with both a DGE and pen-and-paper, use DGEs to discover properties through construction but switch to pen-and-paper to draft and work through the actual proof. 

Another notable category of educational tools are \emph{Intelligent Tutoring Systems} (ITS). While DGEs help students see what is true through manipulation, ITS provide hints and feedback to help students prove why it is true using formal logic. These tools accelerate the learning process by offering instant feedback; this eliminates the wait for manual grading and allows students to fit more practice opportunities into a single session~\cite{koedinger2023astonishing}. Multiple studies on the use of ITS for geometric proof have demonstrated a positive effect on student learning~\cite{campbell2020technology}. \textsc{The Geometry Tutor}~\cite{anderson1985geometry}, an ITS based on cognitive science principles, improved student scores on tests by one standard deviation when compared to their peers who were taught by the same instructor~\cite{anderson1995cognitive}. The quality of feedback provided by an ITS is imperative to the student's success. \citet{paneque2017ggtutor} found that students thought strategically about the proof if feedback was provided that maintained the challenge of the problem. Most ITS rely on experts or teachers to create the feedback that supports the student without giving away the answer. Both DGE and ITS are covered in more detail in \cref{sec:results-rq2} and \cref{sec:tools}.

\section{Interview Study}\label{sec:method}
To answer research questions \ref{rq:1} and \ref{rq:2}, we conducted semi-structured interviews with 18 geometry teachers in the United States. This study received exempt status from the Institutional Review Board (IRB) of Carnegie Mellon University. This exemption was granted because the study consisted of low-risk activities such as anonymized surveys and interviews. Verbal informed consent was collected and recorded from all individual participants included in this study prior to participation, in accordance with the protocol approved by the IRB. Each session lasted between 50 and 80 minutes and was conducted remotely using Zoom. Interviews were led by one researcher, while a second researcher took notes. All audio, video, and screen shares were recorded with participant consent. The interview protocol was organized into three chronological phases, designed to transition from high-level pedagogical goals to situated technological practices, and finally to a concrete problem-solving scenario.

In the first phase, the session began by establishing the participants' professional background and teaching context. The discussion then moved to their philosophy regarding geometric proofs (targeting \ref{rq:1}). We intentionally sequenced the protocol to discuss teaching strategies \textit{before} introducing the topic of technology. This ordering aimed to mitigate priming effects~\cite{strack1992order}, allowing us to capture the participants' pedagogical goals independent of the specific constraints or affordances of their current digital tools.

The second phase narrowed the discussion to specific technological interventions and interactivity (targeting \ref{rq:2}). To understand the participants' situated practices, we employed elements of contextual inquiry~\cite{raven1996using}. Participants were encouraged to demonstrate their actual workflows by screen-sharing existing curricular materials, including digital handouts, interactive modules, problem sets, and online resources. Participants were explicitly asked to share materials they currently use in their classrooms; they did not draw, author, or generate new content during the interview. Only teaching materials, such as answer keys or empty assignments, were shared to avoid revealing student data.

In the final phase, the researcher shared a depiction of the proof seen in \cref{fig:ex-proof} with the participant. Using an artifact elicitation approach~\cite{douglas2015artifact}, we asked teachers to walk through their specific instructional strategies for this problem. This common stimulus grounded the conversation, encouraging participants to speak concretely about the scaffolding tactics they employ and the limitations they face with existing tools.

The interviewer allowed participants to deviate from the procedure to bring up any related topics that came to mind throughout the interview, and some questions were skipped depending on the flow of the conversation. While the semi-structured nature of the interviews allowed for variation, the core instrument guiding these phases was as follows:

\noindent\textbf{Background}
\begin{enumerate}[topsep=.1em]
    \item How long have you been teaching geometry?
    \item What kind of school do you teach at?
    \item What level of geometry do you teach?
\end{enumerate}
\textbf{Teaching geometric proof (targeting \ref{rq:1})}
\begin{enumerate}[topsep=.1em]
    \item How much do you focus on teaching geometric proof in your classroom?
    \item What do you think the value of learning proof is for students?
    \item Walk me through your procedure for teaching geometric proof. 
    \item Do you mostly use textbook problems, online resources, handouts, etc.?
    \item What parts of geometric proof are easiest for your students to understand?
    \item What parts of geometric proof are hardest for your students to understand? How do you explain or target these challenges in your curriculum?
\end{enumerate}
\textbf{Use of technology and interactivity (targeting \ref{rq:2})}
\begin{enumerate}[topsep=.1em]
    \item What types of interactive tools do you use in your classroom? What activities do you use them for?
    \item How often do you use these tools in your classroom?
    \item What do you like about the tools you use?
    \item What challenges or limitations do you run into with these tools?
    \item If no interactive tools are used, why not?
    \item What technology do you use specifically to teach proofs?
    \item Walk me through the last time you used an interactive tool in your class.
\end{enumerate}
\textbf{Artifact elicitation: proof walkthrough}
\begin{enumerate}[topsep=.1em]
    \item How similar is this proof to what your students would be expected to independently complete?
    \item Walk me through how you expect your students to solve this proof.
    \item Which parts of this proof would your students struggle the most with? 
    \item What strategies or activities would you use to help guide them?
\end{enumerate}

\subsection{Recruitment}
Participants were recruited via Facebook posts and email referrals from mutual connections. Subsequently, we used snowball sampling to expand the participant pool to a broader range of teachers. Each participant was compensated with a \$30 Amazon gift card upon completion of their interview.

Teachers were screened for participation based on the following criteria: the participant must (1) be located in the United States, (2) have at least 2 years of experience teaching high school geometry, including geometric proof, (3) have experience creating or adapting curricula for their class, and (4) be willing to share their experience with researchers. Interviews took place between July and September 2024. Our screening survey received 74 responses---more than we could interview---so we used a saturation method to determine the number of participants~\cite{guest2006many}. We achieved a natural saturation point when our interviewees' teaching styles, tools, and insights became consistent with categories that we had already defined, and our analysis stopped revealing new insights. 

The 18 participants come from a variety of teaching backgrounds with an average of 19 years of experience teaching geometry (min of 3, max of 35) at both public and private schools. \cref{tab:participants} shows all of the participants' backgrounds and teaching experiences. 

\begin{table}[ht]
\centering
\begin{tabular}{@{} l l l r @{}}
\toprule
Teacher           & State & School Type &  \thead{Years Teaching\\ Geometry}\\ 
\midrule
P1         & Michigan          & Public     & 21     \\
P2         & Pennsylvania          & Private     & 28     \\
P3         & California          & Private     & 25     \\
P4         & Connecticut          & Public     & 35     \\
P5         & California          & Private     & 18     \\
P6         & Pennsylvania          & Public     & 15     \\
P7         & Pennsylvania          & Private, Catholic     & 26     \\
P8         & Ohio          & Private, Catholic     & 23     \\
P9         & Ohio          & Private, Catholic     & 33     \\
P10         & Pennsylvania          & Public     & 3     \\
P11         & California          & Private     & 25     \\
P12         & Pennsylvania          & Public     & 3     \\
P13         & Ohio          & Private, Catholic     & 5     \\
P14         & California          &  Charter     & 15     \\
P15         & Pennsylvania          & Public     & 13     \\
P16         & Washington          & Public     & 25     \\
P17         & California          & Charter     & 20     \\
P18         & California          & Charter     & 15     \\
\bottomrule
\end{tabular}
\caption{Background of interviewees sorted by participant identifier.}
\label{tab:participants}
\end{table}

\subsection{Qualitative Data Analysis}

All interviews were recorded and transcribed using machine transcription, which was manually checked and corrected by the first author. The third author kept track of high-level insights with notes and affinity diagrams. We periodically reviewed these insights, and when blind spots were identified, the interview protocol was adjusted so those topics could be probed in subsequent interviews. The questions listed above represent the final protocol. 

We analyzed the interview data using inductive thematic analysis~\cite{braun2019reflecting}. The analysis began with an open coding phase, where the authors independently read the transcripts to tag semantic units, such as specific behaviors, pain points, or pedagogical beliefs. These initial codes were collated into themes through a series of collaborative sessions. We iteratively reviewed these themes against the dataset to ensure they accurately reflected the participants' experiences, refining the codebook definitions until theoretical saturation was reached and no new codes emerged. Discrepancies in coding were resolved by revisiting the relevant definitions in the codebook to merge or split ambiguous entries as needed. The results from this analysis are presented in \cref{sec:results} and \cref{sec:results-rq2}.

\section{Interview results targeting RQ1}\label{sec:results}

Our interviews revealed a stark contrast between the abundance of digital tools for general geometry and the scarcity of tools for teaching proof. While participants reported widespread use of \emph{Dynamic Geometry Environments} (DGEs) like \textsc{GeoGebra} and \textsc{Desmos}, nearly all stated they had no adequate software for teaching formal argumentation. As P11 noted, ``So far, I haven't seen any [tools] that felt like they fit particularly well.'' Even teachers with access to premium e-learning platforms expressed frustration with their rigidity.

This section and \cref{sec:results-rq2} unpack this gap between pedagogical needs and technological reality.  Throughout the results, we extract a set of technological requirements that make up the foundation of our review of existing tools in \cref{sec:tools}.  To help practitioners and toolsmiths design future tools, we provide explicit guidance on tool design after every major cluster of study results. The guides connect these results to the pedagogical requirements of the proof learning tools presented in~\cref{tab:reqs}. 

In this section, we focus on the interview results that target \ref{rq:1}. We establish the inherent challenges of teaching proof (\cref{sec:proof-scary}) and the specific skills teachers strive to build (\cref{sec:students-learn}) through their pedagogy. 

\subsection{The pedagogical challenge of teaching proof}
\label{sec:proof-scary}
Teachers note that proof is the very core of mathematics: ``proof is really everything in mathematics that we know, nothing in mathematics is considered a fact unless it is proven''~(P14). It is also the hardest to teach: ``If you were to ask me, `What is the most difficult thing to teach,' it is teaching proofs''~(P12). The task itself is daunting for several reasons. Students are already apprehensive because proof's reputation precedes it: 

\begin{fquote}[(P16)]Students are already a little scared of proofs. So coming in, they've heard stories\dots{}probably from their parents, like `I couldn't do geometry because of the proofs.'\end{fquote}

\noindent Furthermore, everything about proof is foreign to their students, and the resulting struggles can be discouraging for both the teacher and the student: 

\begin{fquote}[(P5)]Formal proof\dots{}is the exact type of thing that turns a kid who doesn't identify as a math person off of math. That is so not interesting\dots{}Who signs up to be a teacher to turn kids off of math?\end{fquote}

From a pedagogical perspective, proof is typically a student's first introduction to logical thinking. Up until geometry, students are used to ``solving for X''~(P12) and the ``procedural problem-solving''~(P17) in algebra. Instead, proofs demand a ``logical argument''~(P9) that is challenging because, ``students haven't been asked to think that way before''~(P13). The reasoning behind a proof is ``out of order''~(P18) and there can be ``multiple ways to go about structuring an argument''~(P11) when solving a proof:

\begin{fquote}[(P14)] When you're trying to do a proof, you can't memorize every proof or every possible avenue of thought that could happen\dots{}There's a bit of creativity in a proof, and it's hard to teach creativity.\end{fquote}

\noindent Even when students have an intuition about whether a proposition can be proven, just knowing where or how to get started is a major hurdle. For instance, P11 said that their students could ``tell that a parallelogram has opposite sides that look like they're going to be congruent,'' but struggled to ``get rolling'' with ``structuring an argument that supports it step-by-step.''

\noindent A further impediment is that the only valid arguments in proof are the ones that are based on established mathematical facts. The naming of all of the different geometric rules is a pitfall that confuses and loses many students:

\begin{fquote}[(P3)]All these crazy rules\dots{}as soon as you do those with ninth graders, you lose 90\% of the class, and the other 10\% are bored.\end{fquote}

\subsection{Instructional strategies for scaffolding proof}\label{sec:students-learn}
By the end of their geometry course, students are expected to be able to independently fill out assigned proofs. Our participants have various strategies to build this mastery. Most are tactics that the students should refer to when they're unsure about how to proceed in a proof.  Having ``a little bit of a checklist or an algorithm that you try to run through can be helpful to get unstuck''~(P11). We outline five key skills that our participants develop with their students. The first three are core to the proof-solving procedure, while the last two deepen their understanding of what constitutes a proof.

\subsubsection{Naming the objects}\label{sec:student-naming}
As teachers introduce geometric proofs to their classes, students first need to understand how to refer to geometric objects like points, angles, segments, rays, triangles, and so on. Objects in geometric proofs have unique naming conventions that need to be taught. For instance, the same angle could be referred to as $\angle AMC$ or $ \angle CMA$, while triangles could be referred to by any permutation of its three points. This can take some time for students to get used to. While pointing to the construction in \cref{fig:ex-proof}, P8 described the difficulty that some of their students have with labeling:

\begin{fquote}[(P8)]Some of my students actually struggle with writing $\angle DMB$\dots{}I can't tell you the millions of times that we've talked about that, but then I'll have an occasional student that won't understand that $\angle DMB$ is one of the two vertical angles, despite the fact that I've said a million times that the $M$ in the middle has to be the vertex.\end{fquote}

\noindent Even for something as simple as naming, teachers use visual tactics to help students attach meaning to the symbols in a proof. For instance, some teachers tell their students to take a ``trip'' around the angle, ``walking from C to M to A''~(P2) to name their objects and avoid confusion.

\begin{bluebox}
\textbf{For tools to support object naming, students should be able to:} \\
\textit{View geometric diagram (\coderef{DE02})}. Teachers have their students reference the construction while naming objects. Tools should provide a view of the construction during proof.

\textit{Refer to objects by standard naming convention (\coderef{DE06})}. Teachers indicate that students are taught specific naming conventions for referencing geometric objects. Tools should ensure that geometric objects can be referenced using appropriate symbols, with standard ordering conventions.

\textit{Receive feedback on naming (\coderef{DE07})}. Teachers indicate that students incorrectly name geometric objects even after multiple reminders. Tools should provide feedback if a student references an invalid object.
\end{bluebox}

\subsubsection{Collecting facts by marking the diagram}\label{sec:student-marks}
As mentioned in \cref{sec:proof-scary}, students struggle with knowing how to start their proofs. Teachers encourage their students to start by ``writing every single thing''~(P17), including the givens and the final goal of the proof. If they still don't see the ``picture of where you're going''~(P8), they're taught to, ``do something you do know''~(P8), even if they're not sure how it will be used in the proof, or work backward from the goal to the givens~(P14). As the students work, they are taught to annotate their diagrams with tick marks to help keep track of what they know:

\begin{fquote}[(P6)]If there's one thing I would like to stress to the students, it is to mark the picture that's accompanied with the proof because a lot of kids just don't do that.\end{fquote}

All participants reported that they teach their students to mark up their diagrams. For instance, P3 is ``really clear with kids on when you have something, mark it.'' These tick marks help students ``see those relationships''~(P8) established during the proof. Marking the diagram helps students keep track of the information established throughout the proof, and plays a big part in the proof-solving process:

\begin{fquote}[(P9)]I tell them almost from day one that geometry is very visual, that marking the picture is going to help you differentiate between [rules such as] $SAS$ and $ASA$, and which one's which. Mark the picture.\end{fquote}

\begin{bluebox}
\textbf{For tools to support fact collection, students should be able to:} \\
\textit{Add tick marks to construction (\coderef{AN01}).} Teachers indicate that students annotate the construction to keep track of known information. Tools should allow students to add tick marks to their diagrams.
\end{bluebox}

\begin{figure}[h]
    \centering
\includegraphics[width=.7\linewidth]{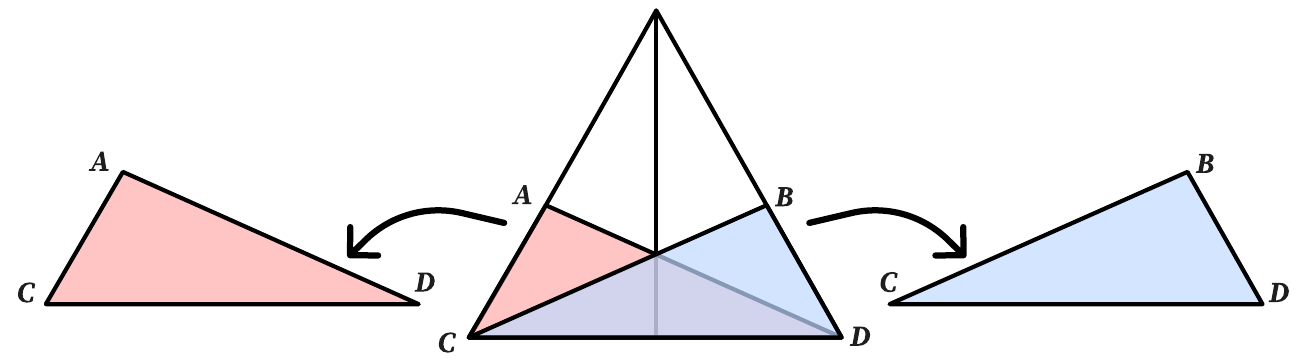}
    \caption{A construction of overlapping triangles $\triangle ACD$ and $\triangle BCD$. Teachers will teach their students to separate out overlapping parts of the construction by redrawing them.}
    \Description{A geometric construction showing two overlapping triangles, labeled as triangle ACD and triangle BCD, sharing side CD. To the left and right of the construction, the triangles are re-drawn separately.}
    \label{fig:overlapping-triangles}
\end{figure}

\subsubsection{Pattern-matching the marked diagram to apply rules}\label{sec:student-pattern}
After collecting information by writing down all of the givens and marking the diagram, the next tactic that teachers emphasize to their students is pattern-matching. Since every step in a proof must be based on previously established information, students learn to use the current state of the proof as a hint to determine what they can conclude next. Teachers want their students to get in the habit of using ``the diagram as a piece of the proof''~(P3). They instruct their students to refer to the diagram, ``to get them to take a diagram and see where it's going before they just start throwing in proof steps''~(P8).

However, this is by no means a trivial task. It's an ambiguous part of the proof-solving process where the student searches for possible options. Many students struggle to pick out valid patterns from the construction:

\begin{fquote}[(P13)]They can look at two angles across from each other on a vertex and recognize they're congruent, but not recognize that they're vertical angles, and they need to use the Vertical Angle Theorem to show their congruence.\end{fquote}

\noindent Teaching students to pattern-match requires a lot of scaffolding and support, but two tactics were common among our participants. The first is \textit{visual decomposition}: helping students pick out familiar visual patterns by having them re-draw or annotate parts of the diagram to isolate specific relationships. For instance, in \cref{fig:ex-proof}, the last step of the proof requires students to recognize that $\overline{AB}$ is a transversal (a line that intersects with two or more other lines) that crosses $\overline{AC}$ and $\overline{BD}$. If students are struggling to notice this, P8 would ``temporarily show AC, BD, and AB extended out. That forms a transversal across two parallel lines, so angles A and B are congruent.'' These types of intermediate visualizations are also temporary; once the student recognizes the pattern, they are erased again, ``because they're no longer helping us''~(P8). Also, different patterns are important at different points in the proof. When P4 helps their students at the board, they observe that they are constantly re-drawing or switching colors as the proof progresses, first to highlight the vertical angles, then the triangles, then the transversal.

Teachers apply this same decomposition strategy to manage the complexity of overlapping triangles (\cref{fig:overlapping-triangles}), a configuration that P8 noted, ``drives some students absolutely crazy''~(P8). To isolate the relevant patterns, both P2 and P8 teach their students to ``trace the first triangle and then separate it away, then trace the second triangle and separate it away''~(P8). P4 facilitates this physically using tracing paper: ``I have them trace over one of the triangles and label it. And then I have them actually take the paper, and either rotate it or flip it so they can see that the whole triangle fits.''

The second tactic is to provide a ``bank''~(P12, P15) of geometric rules that the student can use in their proofs. An important part of the pattern-recognition procedure is, ``just knowing the different ammunition''~(P13) that is available to be used within a proof. P11 opts for ``a large sheet of butcher paper'' with images of the common theorems posted in their classroom and maintains a shared document with every theorem they have learned throughout the semester. P18 allows their students to bring a guide of geometric reasons to their final exam. These rules are ultimately used as reasons within a proof to provide justification for each step.

Although teachers instruct their students to rely on the constructions, they have to be careful because students often assume that the diagram is drawn to scale. Students will include statements based on what appears to be true in the construction. P7 teaches their students not to mark something on their construction until they've written it down in the proof:

\begin{fquote}[(P7)]If things look the same, [students] want to just mark them. I always say, you need to mark your picture, but you can't mark the picture until you write [a symbolic step in a proof] down\dots{}and when you write it down, you have to justify.\end{fquote}

\begin{bluebox}
\textbf{For tools to support pattern-matching, students should be able to:} \\ 
\textit{Decompose construction to search for patterns (\coderef{DE09}).} Teachers have students re-draw or highlight specific parts of the construction to see relationships between geometric objects. Tools should help students temporarily decompose the construction in different ways as they work through the proof. 

\textit{Receive hints about patterns to look for (\coderef{DE08}).} Teachers indicate that students often struggle to pick out visual patterns by themselves. Tools can illustrate these patterns as hints if students are stuck.

\textit{Look up the definition of a reason (\coderef{DE10}).} Teachers provide their students with a glossary of valid geometric reasons. Tools should provide definitions for each geometric reason that a student applies during their proof.

\textit{Provide justification for a proof step with a reason (\coderef{DE03}).} Teachers expect their students to justify each step of the proof with a valid geometric reason. Tools should expect students to use only given and known geometric reasons.

\textit{Receive feedback on tick marks (\coderef{AN02}).} Teachers want their students to mark known information on their diagrams. Tools should provide feedback to students when they mark information that has not been established in the proof.
\end{bluebox}

\begin{figure*}
    \centering
    \includegraphics[width=.9\textwidth]{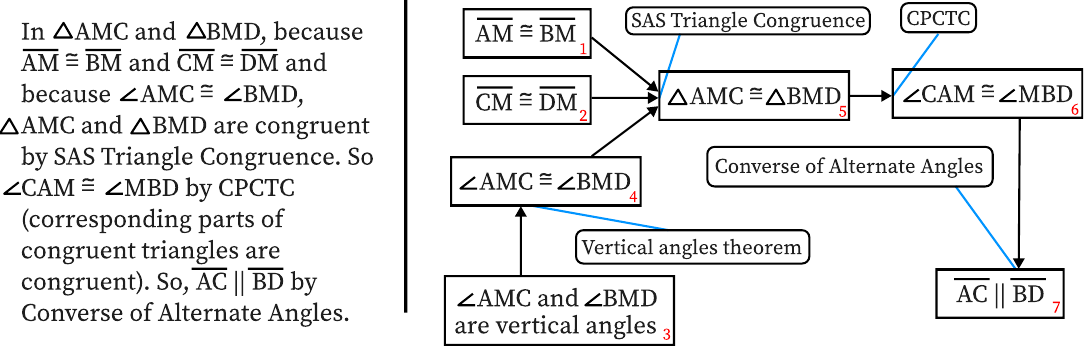}
    \caption{The proof introduced in \cref{fig:ex-proof} illustrated in paragraph format on the \textbf{left} and flowchart format on the \textbf{right}.}
    \Description{A comparison of two proof formats. On the left, the proof is written in paragraph format, presenting the logical steps in a continuous, narrative form. On the right, the same proof is shown in a flowchart format, where each step is represented as a box connected by arrows, illustrating the logical progression of the proof. Note that, since steps 1, 2, and 3 have no incoming connections, they can be listed in any order. Similarly, as long as step 3 appears before step 4, steps 1, 2, and 4 are also interchangeable.}
    \label{fig:ex-paraflow}
\end{figure*}

\subsubsection{Exploring multiple solutions by alternate proof representations}\label{sec:student-multisolution}

While students are only expected to find one solution, teachers want their students to understand that many proofs can be solved in different ways. For instance, in \cref{fig:ex-proof}, steps 1, 2, and 3 are interchangeable, and step 4 could appear before steps 2 or 3. Students often struggle to accept that a proof can have multiple valid structures, as their prior math experience conditions them to expect a single correct solution: 

\begin{fquote}[(P11)]If [students] were writing something for their English or history class, they would never feel like they needed to put the sentences in the exact same order, or use the exact same words as a classmate\dots{}but for some reason around math proofs, students sometimes carry the expectation that it must be that this `line needs to come first, and then this line.'\end{fquote}

\noindent For some teachers, the ``collaborative element''~(P1) of in-class group activities helps to combat the ``pressure that, if this person got a complete proof, that must be the right way''~(P11), and naturally exposes their students to different problem-solving approaches. However,  one major obstacle that obstructs students from recognizing multiple solutions is the proof format itself. Most proofs students encounter in textbooks and on standardized tests are presented in the two-column format. While ``very concise and easier to grade''~(P17), this format has the drawback of being deceptively linear, flattening the actual, underlying tree-like dependency structure between proof steps:

\begin{fquote}[(P13)]
    95\% of the time, we did two-column [proofs] in state testing. All the resource materials I've seen have been two-column proofs, like students are just not asked to construct other types\dots{}Maybe that would help with their thinking, if they did flow proofs, where it doesn't have to be quite as linear and sequenced as a two-column proof.
\end{fquote}

To help students conceptualize or visualize this dependency structure, some teachers introduce proofs in alternative formats, such as flowchart (\cref{fig:ex-paraflow}, right), or paragraph proofs (\cref{fig:ex-paraflow}, left). Some teachers find flow proofs to be more accessible for their students~(P8, P18), because, ``the logical connections are very, very obvious''~(P8). Others find paragraph proofs resonate with their students ``who really like to write''~(P17). In any case, introducing students to multiple formats can help to connect to students who, ``think differently and write differently''~(P8).

\begin{bluebox}
\textbf{For tools to support exploration of multiple solutions, students should be able to:} \\ 
\textit{View two-column proof (\coderef{DE01}).} Teachers indicate that two-column proof is the most common format, and their students need to be familiar with it. Tools should present proofs in the two-column format.

\textit{View proofs in multiple formats (\coderef{FL01}).} Teachers indicate that assigning proof in paragraph or flowchart form can help their students learn that proofs can have multiple solutions. Tools should present proofs in paragraph and/or flowchart format.

\end{bluebox}

\subsubsection{Experimenting with compass-ruler constructions}\label{sec:student-construction}
While the strategies discussed so far allow students to solve standard proofs with provided diagrams, compass-and-ruler constructions represent a distinct skill set with two divergent applications in the classroom. In most contexts, teachers treat construction as a creative, exploratory unit separate from formal logic, using it to build intuition without the pressure of proof. However, in accelerated courses, construction becomes an integral, rigorous step in the proof process itself.

P14 and P11 are the only participants whose students reach this level, expecting them to ``do a geometric construction and then prove it using the theorems we learn in the class''~(P14). These problems become even more challenging when they are under-specified, requiring students to insert \textbf{auxiliary lines}---augmented parts of the construction not explicitly mentioned in the prompt---to bridge the logical gap. These are ``very challenging problems''~(P14) that take groups of students several days to solve:

\begin{fquote}[(P11)]
    I feel like my charge as a geometry teacher\dots{}is to push the students in how they think about math and build skills to think more like a mathematician\dots{}I want to try to destigmatize the idea of being stuck. I want them to hear that mathematicians are stuck all the time, and actually spend more time being stuck than being unstuck.
\end{fquote}

\noindent A lot of the discussion in these classes revolves around hypothesizing different ways to make the construction: 

\begin{fquote}[(P14)]
    I want them to talk to each other about how they're doing the construction and how they know it's correct. And so the conversations in the room are usually\dots{}how do you know that you constructed it? Then they have to defend their construction to their peer.
\end{fquote}

\noindent For these situations, dynamic geometry tools like \textsc{GeoGebra} are strongly recommended to the students to help them hypothesize and iterate on their constructions:

\begin{fquote}[(P14)]
    \textsc{GeoGebra}\dots{}that's a tool I use to help students with the construction, but it doesn't relate too much to proofs, I guess. But it is a way that you can, in an easier way than working with a compass, work out a diagram and experiment with, what if I draw a line here? What if I introduce a circle here?
\end{fquote}

\begin{bluebox}
\textbf{For tools to support experimentation with compass-ruler constructions, students should be able to:} \\
\textit{Create compass-ruler construction (\coderef{EX02}).} Teachers need their students to be able to create constructions. Tools should allow students to build and edit their own compass-ruler constructions.

\textit{Manually test hypothesis (\coderef{EX03}).} Teachers indicate that students must prototype and test their hypothesized constructions before starting on the proof. Tools should allow students to manipulate their constructions.

\textit{Receive feedback on hypothesis correctness (\coderef{EX04}).} Teachers have students discuss their hypothesized constructions so they can receive feedback from each other. Tools should provide feedback on the correctness of a hypothesized construction without necessarily giving away the answer.

\end{bluebox}
    
\section{Interview results targeting RQ2}\label{sec:results-rq2}

We now discuss the interview results that address \ref{rq:2}. We begin by examining how resource constraints shape tool selection (\cref{sec:teacher-pick-tool}). We then detail the use-cases, benefits, and shortcomings of the tools our teachers use, including \emph{Dynamic Geometry Environments} (DGE) in \cref{sec:dge}, and \emph{Intelligent Tutoring Systems} (ITS) in \cref{sec:its-fall-short}. Finally, we discuss the critical lack of integration between diagrams and logical text in current software (\cref{sec:forget-diagram}). Once again, each section ends with the technological requirements derived from its findings, which we revisit in \cref{sec:tools}. 

\subsection{Resource constraints and the reliance on general-purpose tools}\label{sec:teacher-pick-tool}

Geometry teachers are busy. Between running their classes and grading assignments, most are already too busy to recreate their course content year after year. When it comes to proof, very few can take the time to come up with their own problems, so they adapt external resources such as textbooks or online websites to build their materials.

\begin{fquote}[(P15)]
    I haven't created anything. I just go [online] and I find ones that work. And then I can assign them to the students, and I can see what they're doing on my screen.
\end{fquote}

\noindent While teachers curate content for their course, they are also constantly balancing several external factors, including requirements set by their department, school, and government, as well as the individual needs of their students. Additionally, for several teachers ($N=8$) proof ``is only really a target for one unit''~(P12), or a few weeks of dedicated class time, after which it may be revisited intermittently throughout the rest of the school year. So, while proof is the most challenging topic to teach, students also have limited opportunities to master the necessary skills.

Because teachers have limited time to experiment, they tend to rely on general-purpose classroom tools rather than seeking out niche software. All of our participants already incorporate technology into their classes in various ways. A simple example is the use of quiz games like \textsc{Kahoot}~\cite{kahoot} and \textsc{Blooket}~\cite{blooket}, which enable teachers to collect immediate feedback on student misconceptions. Despite a willingness to incorporate technology into their classrooms, many of our teachers ($N=10$) do not use any proof-specific software tools. From P13's perspective, the advancement and availability of proof-specific tools have lagged behind other categories like DGEs:

\begin{fquote}[(P13)]
    We've progressed from giving them paper diagrams to putting them on a PDF and sharing it in \textsc{Google Classroom} or \textsc{OneNote}. That's about the only technology progression that's happened in the last decade.
\end{fquote}

\begin{bluebox}
\textbf{For tools to mitigate resource constraints, teachers should be able to:} \\ 
\textit{Select problems from a bank (\coderef{CU01}).} Teachers indicate that they have limited time to pick up a new resource, and prefer that it provides problems that already work. Tools should provide a bank of proof problems for teachers to choose from.
\end{bluebox}

\subsection{The most commonly used software are DGEs for hypothesis testing}\label{sec:dge}
All participants report using DGEs, such as \textsc{Desmos}, \textsc{GeoGebra}, or \textsc{Geometer's Sketchpad}. However, they were careful to specify that DGEs are considered ``proof-adjacent''~(P10). This is because these tools lack the formal rigor required for proof. DGEs rely on visual measurements to verify properties for specific instances, whereas formal proofs must demonstrate logical validity for all possible configurations of a construction: 

\begin{fquote}[(P14)]
    It's good for maybe discovering a fact, but for establishing it as a fact, that requires a proof.
\end{fquote}

\noindent Instead, these tools are primarily used for induction and discovery. Teachers like that there's a bit of ``playfulness''~(P11) to them, where students can add and move objects around to ``see what happens''~(P2) or ``explore a fact''~(P5). DGEs are a great tool to build intuition and test hypotheses about a diagram before transitioning to a full proof.

\begin{bluebox}
\textbf{For tools to support hypothesis testing, students should be able to:} \\ 
\textit{Create and manipulate geometric construction (\coderef{EX01}).} Teachers use induction and exploration to discover facts and motivate proofs. Tools should allow students to manipulate their constructions.  
\end{bluebox}

\subsection{Teacher perception of their e-learning tools}\label{sec:its-fall-short}

Some teachers’ ($N=8$) schools use interactive e-learning software such as \textsc{DeltaMath}~\cite{korzyk_deltamath_2026}, \textsc{IXL}~\cite{ixl_learning}, \textsc{ALEKS}~\cite{falmagne2003aleks}, and \textsc{Savvas}~\cite{savvas_envision} for individual practice. Using these tools, students solve proofs by filling in blanks, using dropdown menus, or drag-and-drop mechanics. Teachers who have access to these e-learning tools use them more for independent practice rather than synchronous in-class activities.  Each tool has its own unique set of features and constraints. The ITS---\textsc{DeltaMath}, \textsc{IXL}, and \textsc{ALEKS}---promise automatic feedback and personalized learning, while \textsc{Savvas} promises high-quality interactive learning materials. However, according to our participants, the proof-specific activities in each of these four tools fall short of expectations.

\subsubsection{Inconsistencies in the quality of automated feedback}\label{sec:its-feedback}

Each of the four tools provides some amount of feedback to the student. At a minimum, students will be notified if they submit a correct or incorrect answer. Teachers liked the immediate feedback because it allows students to proceed through content at their own pace, without having to wait for the teacher to check their work. The quality of the feedback provided by these tools varies widely. For example, during a walkthrough of \textsc{Savvas}, P8 found some hints more useful than others. For one problem, the only assistance offered was a ``view textbook'' button that ``literally takes you to a PDF of a section of the textbook. It's completely useless''~(P8). On the other hand, the hint for the next problem included a ``six-step scaffolded process to answer the question''~(P8). P8 liked that the student would be forced to do a similar question on their own once they completed the walkthrough.

Teachers appreciated the tools that provided personalized content to their students. For example, teachers who use ALEKS like the ``adaptive knowledge checks''~(P15), where the student takes a placement test, and the program automatically scales the difficulty of the problems to match their skill level, ``individualizing it a little more''~(P13):

\begin{fquote}[(P12)]
    It will create a personalized pathway for the students. So every kid is entering at a different pace\dots{}You might get topic A to start\dots{}Eventually, they give you another learning check\dots{}that recalibrates your learning path. So the goal was to work through all of the topics.
\end{fquote}

\begin{bluebox}
\textbf{For tools to provide automatic support, students should be able to:} \\  
\textit{Receive corrective feedback (\coderef{FE01}).} Teachers indicate that immediate feedback helps their students proceed at their own pace. Tools should, at a minimum, provide feedback about whether a student submitted a correct or incorrect answer.

\textit{Receive explanatory feedback (\coderef{FE02}).} Teachers indicate that explaining the cause of an error or providing worked examples is beneficial for their students. Tools should provide clear explanations for why an answer is incorrect. In cases where worked examples are provided, tools should expect the student to complete a similar follow-up question.

\textit{Receive personalized feedback (\coderef{FE03}).} Teachers indicate that tools that cater to the student's current skill level are desirable. Tools should provide personalized feedback that takes the student's current strengths and weaknesses into account.
\end{bluebox}

\begin{figure}[t]
    \centering
    \includegraphics[width=\linewidth]{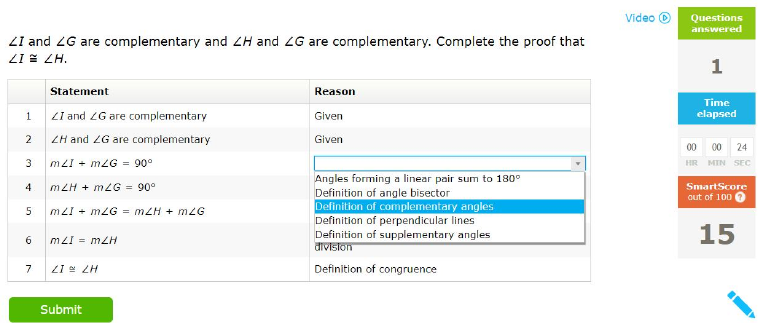}
    \caption{Example of a two-column proof activity in IXL Learning. Students fill in missing information by selecting the correct options from dropdown menus. This proof does not include a construction. On the top right, there is a SmartScore that updates based on the time students spend and the number of questions they answer.}
    \Description{A screenshot of a two-column proof activity in IXL Learning. Students fill in missing statements and reasons by choosing options from dropdown menus. The proof does not include a construction diagram. In the top right corner, there is a SmartScore, which updates based on the time spent and the number of questions answered.}
    \label{fig:IXL}
\end{figure}

\subsubsection{Partial proofs give away the answer}\label{sec:its-exercises}
When students are first introduced to proof, teachers will scaffold activities by providing \textit{partial} proofs. These proofs are mostly complete, and students are responsible for filling in any empty statements or reasons. At first, students are responsible for adding only one statement, but over time, the teacher assigns exercises to fill in all the statements or reasons.

The vast majority of proof exercises in these e-learning tools are partial proofs. Teachers did not like these types of exercises when implemented in e-learning tools, despite regularly using them in their curriculum. The difference between doing a partial proof on paper and in e-learning tools is the \textit{guessability} of the problem. For instance, \cref{fig:IXL} shows a scaffolded two-column proof about complementary angles in \textsc{IXL}. The proof is complete except for the reason of step 3. Because students are allowed multiple attempts and there are only 4 options, students don't actually need to have read the proof; they can look at statement 3 and use an ``elimination method to figure out what's the right step to fill in the reason''~(P13). This can tempt some students into ``button mashing and trying to get through the assignment''~(P13):

\begin{fquote}[(P8)]
    They're not teaching students to take that diagram and have a think-aloud with that diagram and ask themselves before they start writing the proof. What do I know? What can I add to this diagram?\dots{}So I didn't feel that it did a good job teaching students process.
\end{fquote}

\noindent However, there are tool design tactics that can help to discourage random guessing. For instance, \textsc{DeltaMath} takes a simple approach by making the list of possible answers very long:

\begin{fquote}[(P8)]In \textsc{Savvas} I clicked on something and it gave me four choices and Vertical Angles was the only one that any reasonable human would have chosen at that point. Whereas \textsc{DeltaMath} gives you\dots{}lots of different things that you could put in.
\end{fquote}

\begin{bluebox}
\textbf{For tools to support partial proofs, students should be able to:} \\ 
\textit{Complete partial proof (\coderef{DE04}).} Teachers scaffold proof by having students complete proofs with missing information. Tools should provide partial proofs with varying levels of completeness to students. Special care should be taken to ensure that answers are not easy to guess.

\end{bluebox}

\subsubsection{Full proofs need to support multiple solutions}\label{sec:its-full-proof}
While partial proofs are useful when first learning, the ultimate goal is for students to transition to solving full proofs independently. Because full proofs often allow for multiple valid solutions, students must be given the space to experiment with different strategies. Unfortunately, in \textsc{IXL}, \textsc{Savvas}, \textsc{ALEKS}, and \textsc{DeltaMath}, even the most challenging partial proofs are designed to accept only one correct answer. This inadvertently trains students to believe that proofs only have one solution, undermining pedagogical efforts (\cref{sec:student-multisolution}) to demonstrate that proofs can be approached in different ways:

\begin{fquote}[(P11)]
    Every successful proof doesn't have to be exactly the same. Step three doesn't have to be the same in every correct proof. There should be\dots{}space for multiple correct approaches.\dots{}In our pre-calc course\dots{}if [the student's] final answer was correct, that's a pretty solid quick check that [they] could use at home or for extra practice. Whereas for a geometry proof, we know the conclusion when we start. So it's more about the way you get there.
\end{fquote}

Of the four tools, only \textsc{DeltaMath} and \textsc{ALEKS} include activities where students fill out a full proof. These tools provide space for multiple approaches by incorporating a library of pre-programmed valid solutions for each problem.

\begin{bluebox}
\textbf{For tools to support full proofs, students should be able to:} \\ 
\textit{Take multiple approaches through a proof (\coderef{FL02}).} Teachers say that the same proof can often be solved multiple ways. Tools should accept all valid paths that the student can take through the proof.

\textit{Complete full proof (\coderef{DE05}).} Teachers expect their students to build up to completing full proofs. Tools should allow students to solve proofs from beginning to end, proceeding from the given statement to the goal of the proof.

\end{bluebox}



\begin{figure}[b]
    \centering
    \includegraphics[width=.8\textwidth]{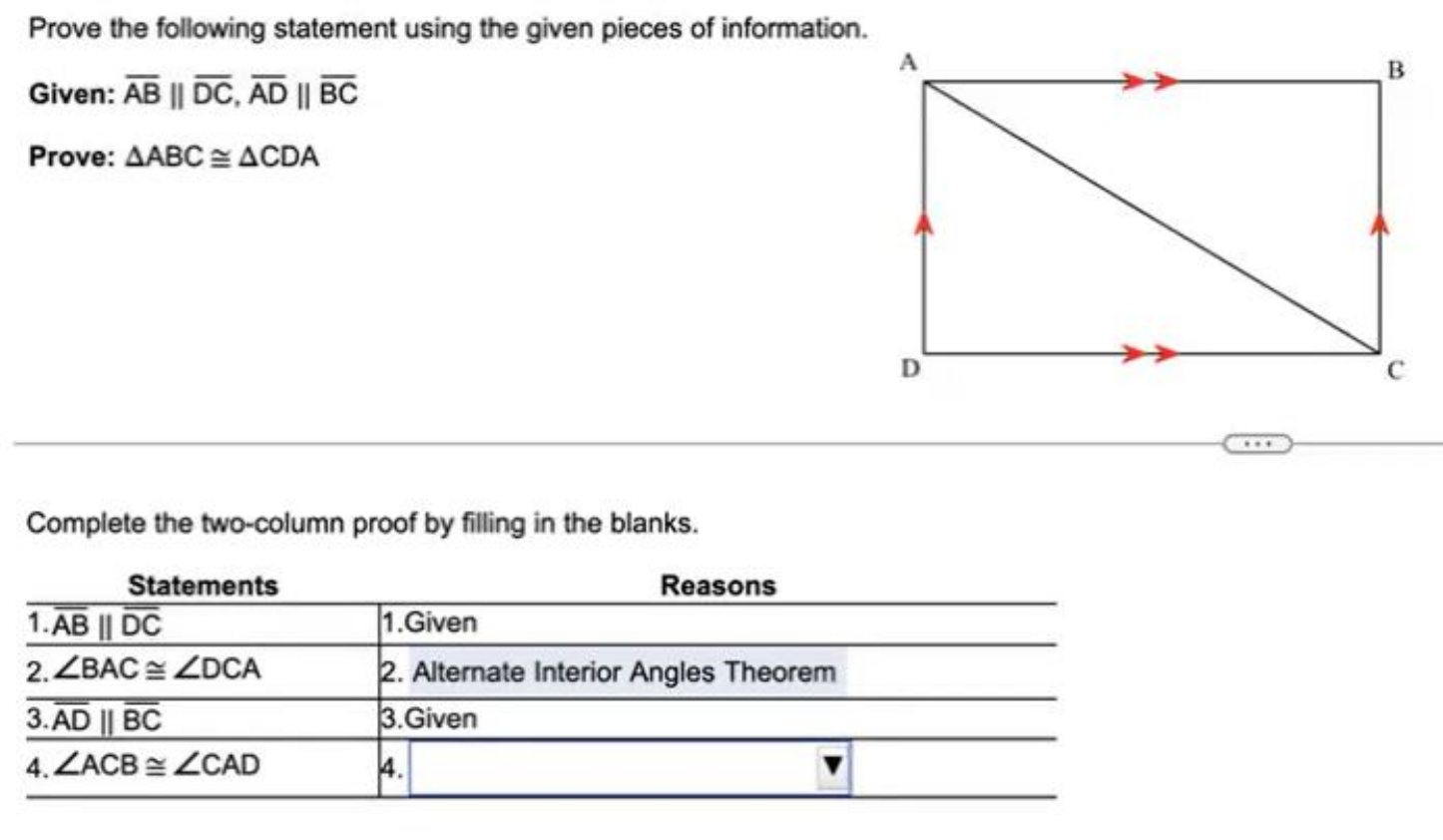}
    \caption{Example of a two-column proof activity in Savvas. Students fill in the empty rows by selecting from a dropdown. The construction is a static image marked with the given information; there is no way for students to interact with or update it as they progress.}
    \Description{A screenshot of a two-column proof activity in Savvas. Students complete the proof by selecting answers from dropdown menus to fill in empty rows. The construction is shown as a static image with given information, and there is no option for students to interact with or update the diagram as they progress through the proof.}
    \label{fig:Savvas}
\end{figure}

\subsubsection{Idiosyncrasies in full proof tools}

 The tools supporting full proofs, \textsc{DeltaMath} and \textsc{ALEKS}, contained various idiosyncrasies that prevented teachers from recommending them enthusiastically. Teachers found that these tools frustrated their students because ``they have their own particular quirks that the students are expected to learn''~(P11). Many of these idiosyncrasies are \textbf{usability} issues that likely originated from assumptions embedded into the tool's design. These add additional hurdles for the students to clear, which are \textit{unrelated} to the core task of doing proof. 

One such example is the naming convention of triangles in \textsc{DeltaMath}. This tool does not follow the same naming conventions as our teachers. Normally, students are taught that triangles can be referred to in any permutation of their three corners. When \textit{classifying} a triangle in \textsc{DeltaMath}, e.g., stating $\triangle ABC$ is isosceles, this naming convention is upheld, and any permutation of the points $A$, $B$, and $C$ is accepted. However, when \textit{comparing} two triangles, e.g., stating $\triangle ABC$ is congruent to $\triangle CEF$, the points must be laid out in a specific order. 

Another example in \textsc{DeltaMath} involves how students add statements to a proof. The student selects an appropriate statement ``template'' from a dropdown and fills it out with the correct object names. These templates, shown in \cref{tab:deltamath}, overload variables: $X$ denotes a point in segments like $\overline{XY}$ yet represents an angle in statements like $\angle X$. This semantic ambiguity introduces \textit{accidental} complexity~\cite{brooks1987no}, forcing students to navigate the tool's idiosyncratic notational logic rather than focusing on the \textit{essential} complexity of the geometric argument. For students who are still learning, this inconsistency adds extraneous details to the already complex process of proof. Students must constantly re-evaluate what the type of each variable is as they work. Furthermore, when students make mechanical errors due to this overloading---perhaps by filling out an ``angle'' template as if it were a point---the system simply flags the step as incorrect. This risks discouraging students by conflating a failure of tool navigation with a failure of geometric reasoning.

\begin{table}[ht]
\centering
\begin{tabular}{@{} l l @{}}
\toprule
Template                    &  Type of $X$ \\
\midrule
$\overline{XY} \cong\overline{WZ}$   &  Point   \\
$\angle X\cong\angle Y$             & Angle    \\
$\triangle XYZ\cong\triangle TUV $ & Point  \\
Classify a triangle                   & ---  \\
Classify a quadrilateral              &    ---  \\
$\angle X$ and $\angle Y$ are complementary/supplementary & Angle  \\
$\overline{XY}$ bisects $\angle X$  & Point and Angle  \\
$XY =\frac{1}{2}WZ$ (or $2WZ$)       & Point        \\
$m\angle X = \frac{1}{2}m\angle Y$ (or $2m\angle Y$)   & Angle   \\
\bottomrule
\end{tabular}
\caption{Abbreviated list of template statements (such as the contents of \cref{fig:ex-proof}, \uilabel{4}) available in \textsc{DeltaMath}~\cite{korzyk_deltamath_2026} for geometric proof. Templates have been copied word-for-word from a screen-share of P8's interview session. The ``Type of $X$'' column specifies the intended type of the variable $X$.}
\label{tab:deltamath}
\end{table}

\subsection{Tools lack integration between diagram and proof}\label{sec:forget-diagram}
As discussed in \cref{sec:student-naming,sec:student-marks,sec:student-pattern}, teachers emphasize a process where the proof and the diagram markings co-evolve. According to our participants, this specific workflow is poorly supported by existing software tools, which lack the flexibility to allow students to use diagrammatic annotations for their logical reasoning. For instance, \cref{fig:Savvas} shows a representative proof problem in \textsc{Savvas}. The accompanying construction is a static image with the givens marked in red. When tools offer no way for the student to add marks to the diagram, teachers are forced to devise workarounds:

\begin{fquote}[(P8)]
I tell my students, if you call me over to help you with a problem, and you have not screenshotted that thing from Savvas and marked it up, I'm gonna politely say, I need you to screenshot that, and mark it up, and I'll be back.
\end{fquote}

\noindent Other tools, such as \textsc{IXL} or \textsc{Student Desmos}~\cite{student_desmos}, offer basic annotation capabilities that enable students to freely scribble over a diagram. However, neither tool provides feedback on the correctness of any marks, which leaves the student to self-monitor.

Tools that attempt deeper integration still fall short. \textsc{ALEKS} offers standalone exercises to practice object identification by ``clicking different parts,''~(P15) of the diagram and receiving a hint if they make a mistake. This interactivity is not carried into the proof-solving environment, where students are only given a static diagram. During proof activities, \textsc{DeltaMath} automatically updates the diagram with new marks as each step is submitted. However, teachers want the student to be responsible for adding the marks by explicitly ``clicking on the pieces [of the diagram] that they want tick marks on''~(P15).

Teachers also expressed a desire for more visual feedback and hints to help guide the student through if they were stuck. \textsc{DeltaMath} only provides feedback when students submit incorrect steps, but P8 wants hints incorporated into the diagram itself:

\begin{fquote}[(P8)]
    Maybe a little help button that\dots{}if they clicked on\dots{}it would highlight the two angles that they should be considering\dots{}so those two angles being shown would [prompt the student], `I'm supposed to do something with those two angles, let me look at this again.'
\end{fquote}

\begin{bluebox}
\textbf{For tools to integrate between diagram and proof, students should be able to:} \\
\textit{Keep proof and diagram state synced (\coderef{AN03}).} Teachers need tools that support the \textit{action} of adding marks after each step in the proof. Tools should expect students to add tick marks themselves, rather than automatically updating the diagram on their behalf.

\textit{Receive feedback and hints as visual cues (\coderef{FE04}).} Teachers indicate a desire for tools that provide visual hints to help their students. Tools could provide hints and feedback by highlighting directly over the diagram or proof to draw attention to what students should focus on next.
\end{bluebox}

\section{Existing Tool Review}\label{sec:tools}

\begin{table}[ht]
\centering
\begin{tabular}{@{} l l l r @{}}
\toprule
Code           & Category & Requirement &  Interview Result \\ 
\midrule
\code{EX01} & Exploration & Create and manipulate geometric construction & \cref{sec:dge} \\
\code{EX02} & & Create compass-ruler construction & \cref{sec:student-construction} \\
\code{EX03} & & Manually test hypothesis & \cref{sec:student-construction,sec:dge} \\
\code{EX04} & & Receive feedback on hypothesis correctness & \cref{sec:student-construction} \\
\midrule
\code{AN01} & Annotation & Add tick marks to construction & \cref{sec:student-marks} \\
\code{AN02} & & Receive feedback on tick marks & \cref{sec:student-pattern} \\
\code{AN03} & & Keep proof and diagram state synced & \cref{sec:forget-diagram} \\
\midrule
\code{DE01} & Deduction & View two-column proof & \cref{sec:student-multisolution} \\
\code{DE02} & & View geometric construction & \cref{sec:student-naming,sec:student-marks} \\
\code{DE03} & & Provide justification for a proof step with a reason & \cref{sec:student-pattern} \\
\code{DE04} & & Complete partial proof & \cref{sec:its-exercises} \\
\code{DE05} & & Complete full proof & \cref{sec:its-full-proof} \\
\code{DE06} & & Refer to objects by standard naming convention & \cref{sec:student-naming} \\
\code{DE07} & & Receive feedback on naming & \cref{sec:student-naming} \\
\code{DE08} & & Receive hints about patterns to look for & \cref{sec:student-pattern} \\
\code{DE09} & & Decompose construction to search for patterns & \cref{sec:student-pattern} \\
\code{DE10} & & Look up definition of a reason & \cref{sec:student-pattern} \\
\midrule
\code{FL01} & Flexibility & View proof in multiple formats & \cref{sec:student-multisolution} \\
\code{FL02} & & Take multiple approaches through a proof & \cref{sec:student-multisolution} \\
\midrule
\code{FE01} & Feedback & Receive corrective feedback & \cref{sec:its-feedback} \\
\code{FE02} & & Receive explanatory feedback & \cref{sec:its-feedback} \\
\code{FE03} & & Receive personalized feedback & \cref{sec:its-feedback} \\
\code{FE04} & & Receive feedback, hints as visual cues & \cref{sec:forget-diagram} \\
\midrule
\code{CU01} & Curation & Select problems from a bank* & \cref{sec:teacher-pick-tool} \\
\bottomrule
\end{tabular}
\caption{List of teacher requirements for software tools to support geometric proof, derived from interview results. Each requirement is defined by an action that the target user---either a student or teacher---should be able to do within a proof-specific tool. Groups of requirements are categorized based on pedagogical goal. (*) Teacher-facing requirement.}
\label{tab:reqs}
\end{table}

Our interviews demonstrate that certain tools are valuable for teachers in some use cases. However, teachers continue to rely on pen-and-paper---a tool that enables freeform annotation at the cost of having no immediate feedback and limited scaffolding for students. There are significant gaps between teachers' needs and the features provided by the tools they are familiar with and use. The geometry tools identified by our participants, however, represent only a small subset of the available options. Designing the next generation of geometric proof tools requires a clear understanding of the current technological landscape, so we now address \ref{rq:3}: how well existing software meets the teacher requirements derived from our interviews.

Our comprehensive set of functional requirements, listed in \cref{tab:reqs}, is drawn directly from our interview results from \cref{sec:results,sec:results-rq2}. These requirements are framed as specific actions for the target user---either student or teacher---within a proof-specific tool. Relevant interview data are referenced in the `Interview Results' column to provide context for each entry. Each requirement fits into one of six functional dimensions of proof instruction. These categories span the entire lifecycle of a proof---moving from initial \textbf{exploration} and \textbf{annotation} to formal \textbf{deduction}, while ensuring the process is \textbf{flexible} and supported by meaningful \textbf{feedback} and teacher-led \textbf{curation}. Together, these requirements represent the core functionality for tools to address the pedagogical demands of geometric proof.  Requirements for usability or accessibility are excluded, as it is not possible to comprehensively evaluate each tool's user experience within the scope of this paper. Instead, our high-level takeaways regarding user experience are discussed in \cref{sec:disc-usability}.

\begin{table}
\centering
\resizebox{\linewidth}{!}{
\begin{tabular}{ll|cccc|ccc|cccccccccc|cc|cccc|c}
\textbf{Category} & \textbf{Tool} & \rotheader{\coderef{EX01}: Create, manipulate construc...} & \rotheader{\coderef{EX02}: Create compass-ruler constr...} & \rotheader{\coderef{EX03}: Manually test hypothesis} & \rotheader{\coderef{EX04}: Receive feedback on hypothe...} & \rotheader{\coderef{AN01}: Add tick marks to construction} & \rotheader{\coderef{AN02}: Receive feedback on tick marks} & \rotheader{\coderef{AN03}: Keep proof, diagram state s...} & \rotheader{\coderef{DE01}: View two-column proof} & \rotheader{\coderef{DE02}: View construction} & \rotheader{\coderef{DE03}: Provide justification for proo...} & \rotheader{\coderef{DE04}: Complete partial proof} & \rotheader{\coderef{DE05}: Complete full proof} & \rotheader{\coderef{DE06}: Refer to objects by standard n...} & \rotheader{\coderef{DE07}: Receive feedback on naming} & \rotheader{\coderef{DE08}: Receive hints about patterns...} & \rotheader{\coderef{DE09}: Decompose construction for...} & \rotheader{\coderef{DE10}: Look up definition of a reason} & \rotheader{\coderef{FL01}: View proof in multiple formats} & \rotheader{\coderef{FL02}: Take multiple approaches throu...} & \rotheader{\coderef{FE01}: Receive corrective feedback} & \rotheader{\coderef{FE02}: Receive explanatory feedback} & \rotheader{\coderef{FE03}: Receive personalized feedback} & \rotheader{\coderef{FE04}: Receive feedback, hints as vis...} & \rotheader{\coderef{CU01}: Select problems from a bank} \\
\hline
\rowcolor{Silver}\textbf{Paperlike} & Pen and Paper & \yes & \yes &  &  & \yes &  & \yes & \yes & \yes & \yes & \yes & \yes & \yes &  &  & \yes &  &  & \yes &  &  &  &  &  \\
  & Classkick &  &  &  &  & \yes &  & \yes & \yes & \yes & \yes & \yes & \yes & \yes &  &  & \yes &  &  & \yes &  &  &  &  &  \\
\rowcolor{Silver}          & Google Slides &  &  &  &  & \yes &  & \yes & \yes & \yes & \yes & \yes & \yes & \yes &  &  & \yes &  &  & \yes &  &  &  &  &  \\
\midrule
\textbf{DGE} & GeoGebra & \yes & \yes & \yes &  &  &  &  &  & \yes &  &  &  &  &  &  &  &  &  &  &  &  &  &  &  \\
\rowcolor{Silver}    & Desmos & \yes & \yes & \yes &  &  &  &  &  & \yes &  &  &  &  &  &  &  &  &  &  &  &  &  &  &  \\
    & Geometer's Sketchpad & \yes & \yes & \yes &  &  &  &  &  & \yes &  &  &  &  &  &  &  &  &  &  &  &  &  &  &  \\
\midrule
\rowcolor{Silver} \textbf{DGE + ATP} & GeoGebra ART & \yes & \yes & \yes & \yes &  &  &  &  & \yes &  &  &  &  &  &  &  &  &  &  & \yes &  &  &  &  \\
          & GeoProof & \yes & \yes & \yes & \yes &  &  &  &  & \yes &  &  &  &  &  &  &  &  &  &  & \yes  &  &  &  &  \\
\rowcolor{Silver}          & OK Geometry & \yes & \yes & \yes & \yes &  &  &  &  & \yes &  &  &  &  &  &  &  &  &  &  & \yes &  &  &  &  \\
\midrule
\textbf{ITS} & IXL &  &  &  &  & \yes &  &  & \yes & \partialf & \yes & \yes &  & \yes &  &  &  &  &  &  & \yes & \yes & \yes &  & \yes \\
\rowcolor{Silver}    & ALEKS &  &  &  &  &  &  &  & \yes & \yes & \yes & \yes & \yes & \yes & \yes &  &  &  &  & \yes & \yes & \yes & \yes &  & \yes \\
    & DeltaMath &  &  &  &  & \yes &  &  & \yes & \yes & \yes & \yes & \yes & \yes & \yes &  &  &  &  & \yes & \yes & \yes &  &  & \yes \\
\rowcolor{Silver}    & ANGLE &  &  &  &  & \yes &  &  &  & \yes &  &  & \yes & \yes & \yes & \yes &  & \yes &  & \yes & \yes & \yes & \yes & \yes & \yes \\
    & Geometry Tutor &  &  &  &  &  &  &  &  & \yes & \yes &  & \yes & \yes & \yes &  &  & \yes &  & \yes & \yes & \yes &  & \yes & \yes \\
\rowcolor{Silver}    & Advanced Geometry Tutor &  &  &  &  &  &  &  & \yes & \yes & \yes &  & \yes & \yes & \yes &  &  & \yes &  & \yes & \yes & \yes & \yes &  & \yes \\
    & Chypre (Cyprus) &  & \yes & \yes &  &  &  &  &  & \yes &  &  & \yes & \yes & \yes &  &  &  &  & \yes & \yes &  &  &  & \yes \\
\rowcolor{Silver}    & Mentoniezh &  &  &  &  &  &  &  & \yes &  & \yes &  & \yes & \yes & \yes &  &  &  & \yes &  & \yes &  &  &  &  \\
    & Cabri-DEFI & \yes &  & \yes &  &  &  &  & \yes & \yes & \yes &  & \yes & \yes & \yes & \yes &  &  &  & \yes & \yes &  &  &  & \yes \\
\rowcolor{Silver}    & Cabri-Euclide & \yes &  & \yes &  &  &  &  & \yes & \yes & \yes &  & \yes & \yes & \yes & \yes &  &  & \yes & \yes & \yes & \yes & \yes & \yes & \yes \\
    & Turing & \yes & \yes & \yes &  &  &  &  & \yes & \yes & \yes &  & \yes & \yes & \yes &  &  &  &  & \yes & \yes & \partialf &  &  &  \\
\rowcolor{Silver}    & QED-Tutrix & \yes & \yes & \yes &  &  &  &  & \yes & \yes & \yes &  & \yes & \yes & \yes &  &  &  & \yes & \yes & \yes & \partialf &  &  &  \\
    & Baghera & \yes &  & \yes &  &  &  &  & \yes & \yes & \yes &  & \yes & \yes & \yes &  &  &  &  & \yes & \yes & \yes & \yes & \yes & \yes \\
\rowcolor{Silver}    & Agent-Geom & \yes & \yes & \yes &  &  &  &  &  & \yes &  &  &  & \yes & \yes & \yes &  &  &  & \yes & \yes & \yes &  &  & \yes \\
    & Géométrix & \yes & \yes & \yes &  &  &  &  & \yes & \yes & \yes &  & \yes & \yes & \yes &  &  & \yes &  & \yes & \yes &  &  &  &  \\
\midrule
\textbf{Teacher-} & Savvas &  &  &  &  &  &  &  & \yes & \yes & \yes & \yes &  & \yes &  &  &  &  &  &  & \yes &  &  &  & \yes \\
\rowcolor{Silver} \textbf{Mediated}   & Schoology &  &  &  &  &  &  &  & \partialf & \partialf &  & \partialf &  &  &  &  &  &  &  &  & \yes &  &  &  & \yes \\
   \textbf{Platform} & Student Desmos & \partialf & \partialf & \partialf &  & \partialf &  &  & \partialf & \partialf & \partialf & \partialf &  &  &  &  &  &  &  &  & \partialf &  &  &  & \yes \\
\midrule
\rowcolor{Silver} \textbf{ATP} & AlphaGeometry &  &  &  &  &  &  &  &  & \yes &  &  &  & \yes &  &  &  &  &  & \yes & \yes &  &  &  &  \\
    & LeanGeo &  &  &  &  &  &  &  &  &  &  &  &  & \yes &  &  &  &  &  & \yes & \yes &  &  &  &  \\
\midrule
\rowcolor{Silver}\textbf{LLM} & ChatGPT Tutor &  &  &  &  &  &  &  & \partialf &  & \partialf & \yes & \partialf & \yes & \yes &  &  & \yes &  & \yes & \yes & \yes & \yes &  &  \\
    & GPT-4o Geometry Demo &  &  &  &  & \yes & \yes &  &  & \yes &  &  &  &  &  &  &  &  &  &  & \yes & \yes & \yes &  &  \\
\rowcolor{Silver}    & Gemini Pro &  &  &  &  &  &  &  & \partialf & \partialf & \partialf & \yes & \partialf & \yes & \yes &  &  & \yes & \partialf &  & \yes & \yes & \yes &  &  \\
\midrule
\textbf{Gamified} & Euclidea & \yes & \yes &  &  &  &  &  &  &  &  &  &  &  &  &  &  &  &  &  & \yes &  &  & \yes & \yes \\
\rowcolor{Silver}\textbf{Learning} & Kahoot &  &  &  &  &  &  &  &  &  &  &  &  &  &  &  &  &  &  &  & \yes &  &  &  &  \\
                  & Blooket &  &  &  &  &  &  &  &  &  &  &  &  &  &  &  &  &  &  &  & \yes &  &  &  &  \\
\hline
\end{tabular}
}
\caption{List of all examined commercial and research tools sorted by category. Each tool is evaluated based on each of the requirements from \cref{tab:reqs}. The ``Pen and Paper'' tool captures the functionality of a printed two-column proof worksheet for baseline comparison. (\yes) User can take the required action within the tool using no workarounds. (\partialf) User can take the required action in some problems or activities within the tool, usually depending on the design of an activity. For example, \textsc{Student Desmos}~\cite{student_desmos} activities can be designed as multiple choice, dynamic explorations, free-response, and so on. In some activities, students are able to manipulate a construction, while in others, they are only able to fill in a proof step.}
\label{tab:tool-review}
\end{table}

\subsubsection{Tool selection criteria}
The tools included in \cref{tab:tool-review} are a combination of the tools our participants use and a review of the literature on tools for geometric proof. There are 33 different tools in 8 categories, including \emph{Dynamic Geometry Environments} (DGE), \emph{Intelligent Tutoring Systems} (ITS), \emph{Large Language Models} (LLM), and \emph{Automatic Theorem Provers} (ATP). Paperlike tools include traditional pen-and-paper along with digital tools that emulate the experience of working on paper. DGE + ATP tools are DGEs that use an ATP to evaluate the generality of specific properties within a construction. Teacher-mediated systems focus on classroom orchestration and rely on the instructor for evaluation, whereas ITS use internal models to provide immediate, granular feedback directly to the learner. Gamified learning tools use game mechanics to drive student engagement and motivation for practice tasks. 

In addition to the tools mentioned by our interview participants, several commercial or research tools were identified. In the ITS category, we included the 11 geometric proof tutors surveyed by \citet{tessier2017etude}, the most comprehensive survey of geometry tutors, as well as their own tutor, \textsc{QED-Tutrix}~\cite{leduc2016qed-tutrix}. In the ATP category, we included ATP built specifically for geometry, as the requirements for geometric proof solvers are fundamentally different than their general-purpose counterparts~\cite{minh2025proof}. Several tools in the ITS category, including \textsc{Chypre}~\cite{chypre1992}, \textsc{Cabri-DEFI}~\cite{cabri-defi}, \textsc{Cabri-Euclide}~\cite{cabri1997}, \textsc{Baghera}~\cite{baghera}, \textsc{Advanced Geometry Tutor}~\cite{matsuda2005advanced}, \textsc{QED-Tutrix}, and \textsc{Géométrix}~\cite{gressier_geometrix}, use ATPs to evaluate student solutions, but they are not included in the ATP category to avoid duplication. For all other categories where a systematic literature review was not available, we picked tools based on two conditions: (1) if they support geometric reasoning, and (2) if they include user-facing features or approaches that are not covered by other tools in our set. For example, multiple machine learning models with advanced geometric reasoning capabilities were identified; however, all use a command line interface, generate text outputs of proofs, and do not provide explanatory feedback in the case of failure. \textsc{AlphaGeometry}~\cite{trinh2024solving} and \textsc{LeanGeo}~\cite{song2025leangeo} are both included because they generate two different types of proofs: \textsc{AlphaGeometry} uses a deductive database method while \textsc{LeanGeo} uses Euclidean axioms. Additionally, though not a publicly available tool, we included the OpenAI demonstration of the tutoring capabilities of \textsc{GPT-4o}~\cite{gpt-4o} as it presents a compelling example of the future of personalized, automated learning.

A glance at \cref{tab:tool-review} reveals that no one category or individual tool supports all of the activities. This indicates that there are still blind spots in tool design and a need for integration between approaches. The remainder of this section presents each category and highlights tools that fulfill its specific requirements.

\subsection{Exploration is for motivation and hypothesis formation}\label{sec:tool-exploration}
Exploration activities are typically done before beginning a proof to help the student explore a construction. Our participants found that compass-ruler constructions aided by DGEs (\cref{sec:dge,sec:student-construction}) are an effective medium for exploration. DGEs allow students to quickly manipulate constructions, measure objects, and view a spectrum of possible versions of a construction. These actions help the student observe relationships between objects in the construction. Exploration activities motivate and contextualize the goal of the proof: proving that the observed relationships \textit{always} hold. However, the feedback of these tools for proof is limited. They can only measure the \textit{current} configuration of the diagram (whatever is rendered on the screen). As such, they cannot generalize an observation. 

One method to automatically test observations within DGEs is to integrate an ATP. This has been explored by \textsc{OK Geometry}~\cite{magajna-okgeometry}, \textsc{GeoGebra ART}~\cite{botana2015geogebraart}, and \textsc{GeoProof}~\cite{narboux2007geoproof}. As shown in \cref{tab:tool-review}, all four core requirements for exploration (\coderef{EX01}-\coderef{EX04}) are fulfilled by these three tools. Students can use them to check if specific observations are provable. For example, in \cref{fig:ex-proof}, they might check whether $\overline{AC}$ is always parallel to $\overline{BD}$. This check is passed to the ATP, which tries to generate a proof using the constraints set on the construction by the DGE. So far, the output of these tools is binary: either the ATP succeeds in proving that the queried property holds, or it returns ``not provable.'' Although it does not help the student form a full proof, this method can help them test hypotheses with increased certainty.

\subsection{Annotation is for bookkeeping}\label{sec:tool-annotation}
Annotation activities refer to note-taking or bookkeeping marks that the student adds to a proof (\cref{sec:student-marks}). These annotations help the student visualize the state space of the proof (\cref{sec:student-pattern}). Teachers ask their students to mark their diagrams to make sense of the proof. Our participants found that tools with a built-in scribble function helped replicate the process of annotating on paper (\coderef{AN01}). Unfortunately, only a few tools support scribbling, including \textsc{IXL}, \textsc{Google Slides}~\cite{google_slides}, \textsc{Classkick}~\cite{classkick}, and sometimes \textsc{Student Desmos}. This was a major limitation of tools such as \textsc{Savvas} and \textsc{ALEKS}, which caused teachers to find workaround solutions, like taking a screenshot of the diagram. None of the tools enforce this marking behavior; the few that provide a built-in marking feature make it completely optional.

The closest example of what this experience might look like is the OpenAI demo of \textsc{GPT-4o}~\cite{gpt-4o} tutoring a student through a geometry problem. During the video, the student uses a tablet with \textsc{GPT-4o} and a geometry problem in a split screen configuration. While the model poses questions to the student, the student answers both verbally and by adding marks on top of the problem content with a stylus. The model responds based on both streams of information. This tight integration between problem, marking, and feedback is a promising approach that should be explored by the next generation of proof education tools.

\subsection{Deduction is for proof-solving}\label{sec:tool-deduction}

Deduction activities are what most would typically classify as proof-solving, where the student chains statements and reasons together to create a logical argument, starting from the given information and proceeding to the goal (\coderef{DE01}, \coderef{DE02}, \coderef{DE03}, \coderef{DE04}, \coderef{DE05}). Based on our interview findings, we synthesized the various strategies from \cref{sec:students-learn} into a step-by-step activity that teachers explicitly train their students to perform. This activity is deceptively simple; it is not a linear task completed in one go, but rather a cyclical procedure that requires the student to consider multiple pieces of information. At every step of the proof, the student engages in a cycle that requires the sequential application of three skills: \emph{fact collection}, \emph{visual search}, and \emph{translation}.

\subsubsection{Fact Collection} 
This is the initial step where the student reads the proof givens and any previously established steps, then translates this information onto the visual state of the diagram. This requires both the ability to name geometric objects (\coderef{DE02}, \coderef{DE06}, \coderef{DE07}) and correctly mark the diagram to reflect the current known facts (\coderef{AN01}, \coderef{AN03}). Our participants noted that many students struggle just to begin a proof because they fail at this initial phase. While most ITS and LLMs, such as \textsc{ChatGPT Tutor}~\cite{openai_chatgpt_tutor} and \textsc{Gemini Pro}~\cite{google_gemini_pro_3} support object naming (\coderef{DE07}) via corrective feedback, they overlook the need to \textit{simultaneously} support active annotation practices (\coderef{AN02}) and enforce consistency between the diagram and the currently known facts (\coderef{AN03}). For instance, \textsc{ALEKS} offers isolated naming exercises that prompt students to read an object name and then visually select it on the diagram. Unfortunately, this functionality is not available in its proof activities. Conversely, \textsc{DeltaMath} and \textsc{ANGLE}~\cite{koedinger1990theoretical} automatically annotate the diagram upon a correct input; while efficient, this deprives the student of the active practice necessary to develop the habit of marking the diagram themselves (\coderef{AN02}, \coderef{AN03}).

\subsubsection{Visual Search}\label{sec:tool-pattern}
With the diagram annotated to reflect the current state, the student searches for a visual pattern that corresponds to a known geometric rule, which would enable the next logical step in the proof. This step relies heavily on pattern-matching (\coderef{DE08}, \coderef{DE09}). Teachers often assist this step by using intermediate representations to visualize these patterns (\cref{sec:student-pattern}). From a tooling perspective, \textsc{ANGLE} offers the most explicit support for pattern recognition. Unlike most ITS that prioritize the chaining of statements and reasons, \textsc{ANGLE} emphasizes the visual process expert proof-solvers use to ``see'' a problem. This approach is rooted in the theory that experts solve proofs faster than novices by identifying high-level ``chunks'' of information, allowing them to mentally skip intermediate steps. \textsc{ANGLE} supports pattern recognition by prompting students to chain these chunks together in a flowchart format. However, this results in proofs that lack the same level of detail that teachers expect from their students and misses requirements like \coderef{DE01} and \coderef{DE03}. Other tools, such as \textsc{Cabri-DEFI} and \textsc{Cabri-Euclide} both provide hints by prompting the student to consider ``sub-problems'' within a proof. When the student is stuck, \textsc{Agent-Geom}~\cite{cobo2007agentgeom} sometimes provides hints about how to break down a construction into chunks. Most tools, however, entirely skip this visual search process and move directly to translation. This oversight misses the potential of interactive technology, and future tools should support visual search explicitly.

\subsubsection{Translation} 
Once a visual pattern is identified, the student must translate this understanding back into a formal geometric rule (\coderef{DE03}, \coderef{DE10}), selecting the correct relationship between the correct objects (naming objects, \cref{sec:student-naming})---for example, adding a step to their proof stating that two triangles are congruent by SAS triangle congruence. For most tutors, this is when ITS checks the student's work. However, by only checking the output of this cycle, tools miss opportunities to provide valuable feedback and scaffolding for the preceding two steps. Additionally, because many of our teachers find memorization of geometric reasons to be nonessential to the proving process, they support students by providing definitions of allowed reasons (\coderef{DE10}). Some tools, such as \textsc{ANGLE}, \textsc{Geometry Tutor}, \textsc{Advanced Geometry Tutor}, and \textsc{Géométrix}, fulfill this requirement by providing a glossary for students to refer to. 
Once the student has correctly translated the step, they are expected to update the diagram with the newly concluded information, and the cycle repeats. The effort spent on teaching proof, as emphasized by our interviewees, is largely dedicated to helping students successfully navigate and internalize this three-step deductive cycle. Annotation is thus deeply entangled with deduction; the diagram markings track the current state after every successful cycle. Tools must incorporate support for each major step in this cycle to effectively enforce this best practice.

\subsection{Flexibility reinforces that proofs have multiple solutions}\label{sec:tool-flexibility}

Flexibility is a critical requirement for proof-specific educational tools, ensuring that students practice solving geometric proof beyond rigid, single-solution paths. Two major categories of flexibility emerged from our interview results: support for multiple solution paths and support for multiple proof representations.

\subsubsection{Supporting Multiple Solution Paths}

The nature of proof allows for multiple solutions because students can often choose different correct rules or established facts, creating various valid routes to the final answer. Teachers often use group activities to expose their students to multiple solutions (\cref{sec:student-multisolution}). As discussed in \cref{sec:its-fall-short}, a major limitation of tools that teachers use, such as \textsc{IXL} and \textsc{Savvas}, is that all of their problems related to proof only have one correct answer. To make matters worse, questions are boiled down to simple multiple-choice problems which allow students to brute-force their way through exercises. This approach inadvertently trivializes the genuine ambiguity and complexity inherent in proof construction, which students need to master.

Tools must support multiple proof solutions (\coderef{FL02}) to match the experience of completing a full proof independently. However, this is a technically challenging feature to implement. Tools generally adopt one of two strategies to address this. The first relies on teachers or experts to manually code the correct solutions for every proof problem. This approach is generally not scalable as it requires high effort for each new problem. The more scalable strategy uses automatic deduction to generate the full solution space (the ``proof tree''). This approach has already been explored by several existing systems such as \textsc{Chypre}, \textsc{Baghera}, \textsc{Advanced Geometry Tutor}, \textsc{QED-Tutrix}, and \textsc{Géométrix}.

\subsubsection{Supporting Multiple Proof Representations}
Most proofs students encounter in textbooks and on standardized tests are presented in the two-column format (\cref{sec:student-multisolution}). Although it is the most common, this format is deceptive. It flattens the actual, tree-like dependency structure between proof steps and presents it in a procedural order. This format also leaves no physical space to suggest alternate paths through the proof.

To help students conceptualize or visualize this inherent structure, teachers introduce multiple proof formats (\coderef{FL01}), including paragraph and flow (\cref{sec:student-multisolution}). Software tools can visualize this more efficiently than pen-and-paper methods. Tools like \textsc{QED-Tutrix} and \textsc{Cabri-Euclide} provide some support for multiple representations by having the flowchart and two-column formats synced and placed side-by-side in their interface. \textsc{Mentoniezh}~\cite{mentoniezh} uses a staged approach where the student first uses the two-column format to structure their argument and then translates this outline into a full paragraph proof. Being able to alternate or transition between different proof output formats efficiently helps students understand the structural logic of a proof, instead of viewing it merely as a linear sequence of statements.

\subsection{Feedback helps students work independently}\label{sec:tool-feedback}

The primary advantage software tools have over pen-and-paper methods is the ability to provide immediate feedback on student work. With pen-and-paper, teachers must manually check and provide feedback to each student, which is a time-intensive activity. Software, however, can use the problem state and student interactions to deliver timely guidance.

ITS promise multiple layers of such immediate support, and every tool under this category implements this to some extent. A recurring issue noted by our participants concerning existing industry tools is the inconsistency of feedback across different problem types (\cref{sec:its-feedback}). For example, when students are stuck, some problems might only offer a generic ``try again'' prompt, while others might provide a more helpful, guided walkthrough of a similar example.

Scaling high-quality feedback for proof problems is expensive. Each proof can have multiple solution paths and multiple different types of feedback. Many ITS require an expert or teacher to manually code the feedback associated with specific steps or errors. This overhead significantly limits the scalability of these tools. Expecting teachers to spend time recording pre-planned feedback for every potential error across many new problems is unrealistic given their existing time and resource constraints (\cref{sec:teacher-pick-tool}).

There are two promising technical approaches to scaling the availability of high-quality, auto-generated feedback: ATPs and LLMs. Integrating ATPs into the system could allow the tutor to auto-detect a wider range of errors and misconceptions, thereby providing more detailed, contextual feedback. This approach requires an ATP sophisticated enough to analyze partially incorrect proofs and still generate detailed diagnostic information. Also, the ATP's generated proofs must align with the skill level and expectation of a high school student, as proofs derived using advanced logic would be inaccessible~\cite{font2020automating, minh2025proof}. Two notable examples of this are the ATP incorporated into \textsc{QED-Tutrix}~\cite{font2020automating} and the GRAMY prover~\cite{matsuda2004gramy} as they are specifically designed to match the Euclidean geometric proofs taught at the high school level. However, these provers can only provide binary feedback about the correctness of the proof to the student. Their system still relies on the teacher or hard-coding to determine when and what explanatory or personalized feedback to provide. 

Recent advances in LLMs provide a promising future for highly personalized feedback that can react dynamically to student actions and questions. The demonstration of \textsc{GPT-4o} illustrates how a system could provide interactive prompts, allowing the student to respond through free-form input. This capability more closely replicates the experience of working with a private tutor, offering a potentially powerful and scalable solution for nuanced, on-demand feedback in proof education. However, LLMs can also hallucinate misleading or incorrect information~\cite{llmhallucination2025survey}. One approach that is used in some domain-specific machine learning models, including \textsc{AlphaGeometry}, is establishing a ground truth to ``fact check'' the LLM. In this case, an ATP could check the LLM's output and catch hallucinations before they reach the student. 

The vast majority of tools only provide feedback as text. However, teachers noted that they often mark or highlight errors on the proof or construction. Only a few tools provide any visual feedback or hints to the student (\coderef{FE04}). For visual feedback, \textsc{Geometry Tutor} and \textsc{Baghera} color incorrect inferences in red, while \textsc{Cabri-Euclide} automatically generates a construction of a counterexample when the student submits an incorrect step. \textsc{ANGLE} provides hints by highlighting directly over the construction as the student works. The geometry construction game \textsc{Euclidea}~\cite{euclidea} offers hints by displaying a sequence of tool icons---such as the circle, perpendicular bisector, or segment tools---that represent the necessary construction operations to complete a puzzle. These diverse examples suggest that moving beyond text-based corrections toward integrated, graphical interventions would better mirror teacher practices and transform the diagram from a static reference into a dynamic pedagogical tool that ``speaks'' to students in the visual language of the domain.

\subsection{Curation makes tools accessible to teachers}\label{sec:tool-teacher}
As discussed in \cref{sec:teacher-pick-tool}, teachers have to weigh many factors when designing their curriculum. Except for accelerated courses, dedicated proof instruction typically spans only a few weeks. In reality, most of our participants adapt or select from existing resources that fit their needs. 

Software tools can support teachers by lowering the barrier to adoption. They typically adopt one of two approaches: the first is to provide a sizable bank of problems and activities of varying difficulty (\coderef{CU01}). The activities are either automatically selected by the software or assigned by the teacher. Several tools, including \textsc{IXL}, \textsc{ALEKS}, \textsc{DeltaMath}, and \textsc{Schoology}~\cite{schoology} take this approach. The second approach is to incorporate the teacher into the authoring process. Tools such as \textsc{QED-Tutrix}, \textsc{Turing}~\cite{turing2007}, \textsc{Mentoniezh}, and \textsc{Cabri-Euclide} take this approach. Unfortunately, the amount of data required to make a single flexible proof problem with feedback leaves teachers with an unrealistic workload. Authoring a single problem could involve any or all of the following activities: creating a construction, writing the problem, choosing or authoring the allowable solutions (\cref{sec:tool-flexibility}), and incorporating custom feedback (\cref{sec:tool-feedback}). The cost of authoring a single proof problem must be dramatically reduced for teacher authoring to become a realistic option.

\section{Discussion}\label{sec:discussion}

Based on the needs of geometry teachers and a comprehensive evaluation of existing tools, we can now discuss the gaps and propose research directions for the next generation of proof education tools. We argue that modern technology makes it possible to bridge the gaps identified in our review. Although building a single tool that solves every pedagogical challenge is difficult, we believe there are some ``easy wins'' where specific technical features yield immediate pedagogical payoffs. For instance, implementing ``live,'' semantically connected diagrams supports \coderef{AN01}, \coderef{AN02}, \coderef{AN03}, \coderef{DE08}, \coderef{DE09}, and \coderef{FE04}, while integrating LLM-driven feedback mechanisms can address \coderef{DE07}, \coderef{FE01}, \coderef{FE02}, and \coderef{FE03}. By using the framework established in this paper, tool designers can clearly articulate their value proposition to teachers and evaluate the pedagogical return on investment for these new technical features. However, there are also trade-offs to consider in these implementations. Challenges such as generating problems at scale, providing accurate, personalized feedback, and making interactive activities are non-trivial tasks for the research and industry communities.

Despite these challenges, we believe the best path forward is to combine the strengths of different tools. We establish in \cref{sec:tool-exploration} that exploration activities are already well-supported by DGEs. Tools to support deduction activities, such as ITS, are still falling short of teacher requirements. In particular, they overlook the integration between the diagram and the written proof and rely on content experts or teachers to define the ``automatic'' feedback. The recent advances in LLMs and ATPs, as well as modern interactive web design principles, present an opportunity to combine aspects of each piece of technology into a seamless experience that far outpaces anything in the current space. The following sections outline what we consider to be the critical improvements required for the next generation of proof education tools.


\subsection{Proof tools should have strong visual-symbolic links}\label{sec:disc-interactivity}

The results from \cref{sec:students-learn} reinforce that geometric proof is a visual problem as well as a logical one~\cite{laborde2005hidden}. Typically, proofs are checked by validating the structure of the logical argument. However, what most tools miss are the ungraded, visual tactics that \textit{support} students through the proof-solving process. Our interviewees repeatedly emphasized the importance of several such tactics, including referencing the diagram to correctly name objects (\cref{sec:student-naming}), marking the diagram to keep track of the known information (\cref{sec:student-marks}), or re-drawing components of the diagram to find patterns (\cref{sec:student-pattern}). \cref{sec:tools} demonstrated that certain tools provide support for marking (i.e., scribble tools, \cref{sec:tool-annotation}) or pattern-matching (\cref{sec:tool-pattern}). However, we believe that these tools fall short due to the lack of integration between the diagram, the annotations made by the student, and the proof. For instance, in \textsc{IXL}, scribbles can be added on top of a static image, but there is no way for \textsc{IXL} to check if the student's marks are correct, and no way to enforce that the student actually marked the diagram to begin with.

Building a geometric proof tool with integration between these sources requires the coordination of multiple technical and design components. The first key requirement is to maintain semantic consistency between the diagram and the state of the proof. However, this is difficult to achieve in standard direct-manipulation interfaces, where shapes are typically treated as semantics-free graphics once they are dragged onto the canvas. Lacking an underlying logical representation, these visual elements lose their mathematical identity and cannot automatically update to reflect changes in the proof state.

Tools such as \textsc{Penrose}~\cite{penrose} or \textsc{Mermaid}~\cite{sveidqvist2021official} use text specifications to generate diagrams with a one-to-one mapping between semantic content and visual output. Using a similar approach can allow students to update or interact with the diagram while the system checks for consistency between the diagram and the proof. This could power the development of several pedagogical features, such as dynamically highlighting, adding, or removing portions of a diagram based on the current state of the proof. \citet{head_math_2022} described different augmentations that can be used to aid understanding of math formulas. A similar approach could be adopted to display errors across both the diagram and proof simultaneously. Other features could include automatically verifying that student-added marks match their textual claims, and generating dynamic visual scaffolding based on proof progress.

To fully realize this potential in an educational context, this mapping must be bidirectional, unifying the diagram's state with the underlying proof representation. This synchronization ensures that changes in the proof text can be reflected on the diagram, and conversely, that diagrammatic manipulations can update the proof logic. Such integration unlocks the technical foundation for many of the least-supported requirements, including \coderef{AN01}, \coderef{AN02}, \coderef{AN03}, \coderef{DE08}, \coderef{DE09}, and \coderef{FE04}.  However, implementing this architecture presents a significant technical challenge, as it requires seamless coordination between modern reactive frontend frameworks and a rigorous backend proof checker to maintain logical validity without sacrificing interface responsiveness.  

\subsection{Proof tools should provide personalized problems at scale}\label{sec:disc-proof-checking}

Currently, designing effective proof problems creates a prohibitive ``one-to-many'' content bottleneck: for every single problem, the author must anticipate multiple valid solution paths and prepare specific feedback for every potential error. The requirements defined by our teachers---providing accurate, helpful automated feedback and supporting multiple approaches---exacerbate this challenge. To accomplish this at scale, a tool must track the state and correctness of the proof without requiring manual input for every step. While general theorem provers can theoretically track state, they are often inadequate for the specific pedagogical context of high school geometry (\cref{sec:tool-feedback}).

\begin{figure}[t]
    \centering
    \includegraphics[width=.9\linewidth]{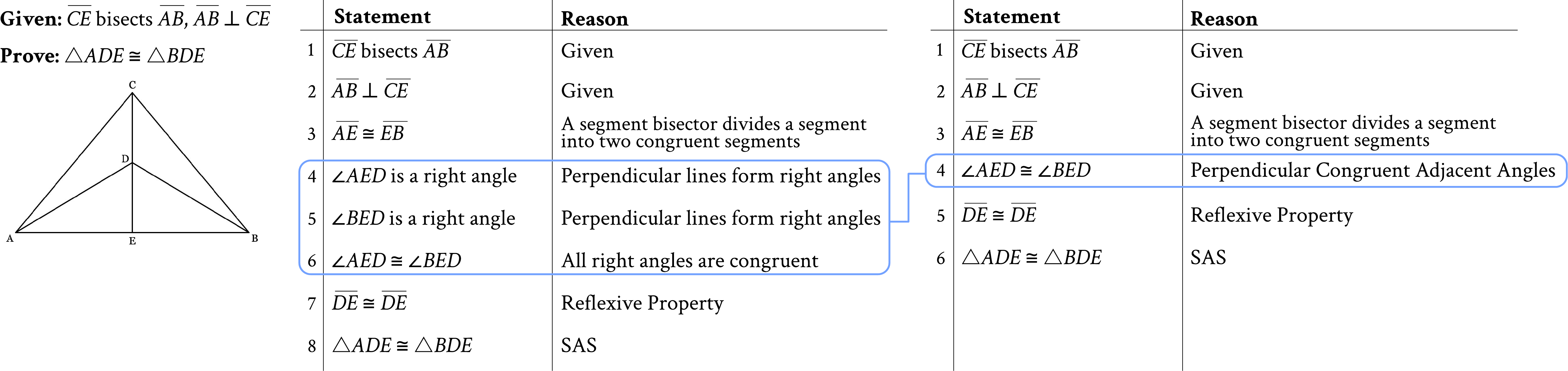}
    \caption{\textbf{Left:} One of the full proof problems available in \textsc{DeltaMath}. \textbf{Center:} \textsc{DeltaMath}'s solution to the proof. Note that the student is expected to explicitly state angles $\angle AED$ (step 4) and $\angle BED$ (step 5) are right angles, before stating that they are congruent in step 6. All three of these steps are necessary in order for \textsc{DeltaMath} to accept step 8. Our teachers said that step 4 and 5 are extraneous, and would confuse their students. \textbf{Right:} Example of teacher's solution to the same proof. Step 4,5, and 6 are combined into one step.}
    \label{fig:deltamath-proof}
\end{figure}

\subsubsection{Automating Problem Generation} To reduce the cost of problem generation, we propose the use of an ATP built specifically for high school-level geometry. \citet{keenan2024learner} detailed a set of requirements for classroom proof assistants. Most of these also apply to the special case of high school-level geometry, including providing an explanation of errors, providing feedback even when errors are encountered, preventing exercises from being completed with guessing or automation, and using notation similar to written proofs. Crucially, we add that the prover should only accept properties available at the high school level. This requirement helps to produce proofs that are understandable to the student and was also identified by \citet{font2020automating}. By adhering to these constraints, a specialized ATP can automatically generate the solution space and corrective feedback, freeing the teacher from tediously defining their own solutions. 

However, there are tradeoffs to consider with this approach. An ATP typically requires a greater level of rigor and detail in its proofs than what students are taught, meaning some logical steps must be abstracted away from the student. If the abstraction does not match the student's mental model, it becomes a point of confusion. For instance, \cref{fig:deltamath-proof} shows an example of a proof in \textsc{DeltaMath}. The two-column proof on the left shows that \textsc{DeltaMath} expects students to declare both angles to be right angles before they can be declared congruent. The proof on the right side shows how students expect to complete the proof, immediately declaring the two angles congruent by observing that they are adjacent angles formed by perpendicular lines. Our interview participants said that these sorts of differences were very frustrating for their students to understand. Although the fix seems straightforward---align the solver with the student's expectation---not every teacher uses exactly the same standards. Some teachers may expect more or less detail depending on the proficiency of their students. \citet{font2020automating} attempt to resolve this by allowing teachers to specify which steps can be omitted, but this approach inevitably shifts the burden back onto the instructor. A comprehensive approach would support solutions that provide different levels of detail without requiring constant teacher intervention.

\subsubsection{Scaling Explanatory and Personalized Feedback} Beyond the structural correctness of the proof, providing meaningful explanatory feedback remains a major hurdle (\coderef{FE02}). As discussed in \cref{sec:tool-feedback}, one major challenge of existing ITS is that they heavily rely on experts to hand-craft feedback for every potential mistake. This places a multiplicative implementation cost on the expert and limits the feedback to exactly what was pre-authored, making personalization difficult (\coderef{FE03}). LLMs may be the solution to addressing this issue. LLMs have successfully been used to generate tutor feedback with the same learning gains as educator feedback~\cite{zhao2026llm}. To ensure that the LLM does not hallucinate improper feedback, one potential solution is to use detailed fault localization from a specialized ATP to fact check the LLM's output. Then, the combined rigor of the ATP and the flexibility of an LLM could be extended to support other forms of assistance, such as generating additional questions, worked examples, or answering unexpected questions on the fly. 

Finally, LLMs can assist in tracking the student's learning trajectory through \emph{Knowledge Tracing} (KT). KT predicts a student's ability to answer the next question correctly based on their past interactions.  Maintaining an accurate understanding of the student's overall progress is crucial for ITS to deliver a personalized learning experience. Traditional KT is labor-intensive, relying on experts to manually model the individual ``knowledge components'' a student must master. Recent research has explored combining KT with LLMs to automatically discover these knowledge components~\cite{wei2025kcluster}. This approach can help to overcome some of the pitfalls of traditional knowledge tracing approaches, though its success currently depends on the availability of structured, limited datasets~\cite{cho2024systematic}. Still, it offers a glimpse into how current breakthroughs in AI could continue to improve the level of personalization that educational tools can achieve.

\subsection{Proof tools should support multiple formats}

The first two improvements combine to reduce the overall effort required to make one proof problem. This benefit can also be used to explore new or different types of content that would otherwise be too costly to implement. For instance, teachers sometimes assign different proof activities in different visual formats, such as flowchart~\cite{miyazaki2014functions} or paragraph, to help their students grasp the underlying logical structure (\cref{sec:student-multisolution}). However, it is uncommon for teachers to represent the \textit{same} proof using multiple formats. Representing the same proof in different ways could help students build \emph{metarepresentational competence}~\cite{disessa_metarepresentation_2004}. Specifically, \citet{rau_framework_2017} showed that students achieve better conceptual understanding through interacting with multiple visual representations. Therefore, tool designers should consider supporting multiple views of the same proof (supporting \coderef{FL01}). The output of a proof-checker (\cref{sec:disc-proof-checking}) could be fed into a system to render the same proof in many different formats. This approach is similar to that of the Lean theorem prover's \textsc{ProofWidgets}~\cite{nawrocki_extensible_2023}, an extension to the Lean interface that allows custom proof visualization. In addition to the standard ``info view,'' a list of the goal and the proof state, \textsc{ProofWidgets} can visualize proofs as customized, domain-specific diagrams (e.g., Euler diagrams for set theory and tree diagrams for tree search algorithms in computer science). This could allow researchers to explore the effect of presenting the same proof in different formats, or even allow students to toggle between multiple representations as they work. 

\subsection{Proof tools should be usable}\label{sec:disc-usability}
In the words of P12, ``remember, proof is going
to be the worst part of [a student's] day.'' Proof requires the student to sustain a high cognitive load while managing various sources of information~\cite{paas2003cognitive, sweller_why_1994}. In other words, the task is \textit{essentially} complex~\cite{brooks1987no}. A meta-issue we observed is that many of the proof-specific tools suffer from various usability issues that introduce \textit{accidental} complexity. Some of these complexities were described in \cref{sec:its-fall-short}, but it is not possible to enumerate every one we encountered within the scope of this paper. Every time the student has to screenshot a diagram and mark it for their teacher, switch to a new window, or search through tabs and dropdown menus, there is a switching cost. These interactions are all sources of accidental complexity for what is already a cognitively demanding task. Some of these costs may be small enough to ignore, but our observation is that most sources of accidental complexity could be resolved with intentional, careful user interface and experience design. Special care should be taken to minimize these sources wherever possible.

Lastly, teachers are resource and time-limited. Most of our participants only have one unit devoted to introducing proof, which is a few weeks of dedicated class time at most. Tools specifically for the learning and doing of proof are desired, but they must be easy to adopt for their classes (\cref{sec:tool-teacher}). Most teachers do not have time or energy to devote to learning a new tool, teaching their students to use it, and adding their own problems. Tools should be careful to provide good default behaviors and a bank of problems for teachers to choose from, and optimize for keeping the cost of onboarding low. If problem-editing capabilities are provided, they should be as easy, if not easier than, creating a printed worksheet. Tools should be easy to set up and access, which can be accomplished by using web-based frameworks.  

\section{Conclusion}\label{sec:conclusion}

Geometric proof remains a challenging and multifaceted topic to teach and learn. Through our investigation of teacher needs (\ref{rq:1}, \ref{rq:2}) and the landscape of existing technology (\ref{rq:3}), we revealed a critical disconnect between the skills teachers strive to build and the digital tools available to support them.

Our interviews with 18 geometry teachers highlighted that successful proof instruction relies on a scaffolded \textit{deductive cycle} consisting of Fact Collection, Visual Search, and Translation. This cycle is inherently visual; teachers rely on diagrammatic annotation and pattern-matching as key aspects for student proof-solving. However, our review of 33 existing tools exposes a segmented workflow: teachers must choose between tools that support annotation and tools that provide immediate feedback. The lack of integration between the \textit{visual state} of the diagram and the \textit{logical state} of the proof remains the largest oversight in the current landscape.

These findings underscore that the challenge of teaching proof is not merely a content problem, but also an interaction design problem. Closing the gap between teacher needs and technological reality requires an interdisciplinary approach that combines the precision of formal methods with the empathy of user-centered design. We believe that the HCI community is uniquely positioned to address this challenge by designing interfaces that reduce accidental complexity and support the non-linear process of student learning. \cref{sec:discussion} outlined necessary improvements for the next generation of proof education tools. In particular, many of the unaddressed requirements derived from our interviews would be resolved by the development of ``live'' diagrams for geometric proof---interfaces where the diagram is both interactive and semantically linked to the formal proof. Reducing the effort required to build proof problems also makes it easier to implement, evaluate, and explore different visual proof formats. Throughout these developments, tool designers must be careful not to introduce accidental complexity into the user experience. Proof is already such a complex topic, so any additional complexity risks overwhelming the student. By aligning algorithmic capabilities with the human realities of the classroom, we can transform geometric proof from a gatekeeping mechanism into an accessible gateway for logical reasoning.

\section{Generative Artificial Intelligence Use Disclosure}
During the preparation of this manuscript, the authors used Gemini 2.5 Pro to convert outlines into initial rough text and to brainstorm alternative phrasing for select sections. All text generated by Gemini was thoroughly reviewed, refined, and replaced as necessary by the authors to ensure accuracy and alignment with the paper's intent. Additionally, Scite~\cite{nicholson2021scite} was used to assist with the literature review process, specifically aiding in the development of \cref{sec:intro} and \cref{sec:related}. To maintain academic integrity, all references provided by Scite were manually validated by the authors prior to their inclusion in the final manuscript.

\section{Funding Disclosure}
This material is based upon work supported by the National Science Foundation under Award No(s): 2447499, 2346174, and 2119007. It is also supported by the National Science Foundation Graduate Research Fellowship Program under Grant No(s) DGE2140739. Any opinions, findings, and conclusions or recommendations expressed in this material are those of the author(s) and do not necessarily reflect the views of the National Science Foundation. Funding was also received from the ARCS (Achievement Rewards for College Scientists) Foundation, Pittsburgh Chapter.

\section{Disclosure Statement}
The authors have no financial or non-financial competing interests to report.

\section{Author Contributions}
Hwei-Shin Harriman was involved in the conception and design of the study, the analysis and interpretation of the data, and drafting the paper and revising it critically for intellectual content. Wode Ni was responsible for the interpretation of the data, drafting the paper, and revising it critically for intellectual content. Yuchen Jin was responsible for the design of the study and the analysis of the data. Dominik Moritz and Joshua Sunshine were involved in the conception of the study, the interpretation of the data, revising the paper critically for intellectual content. All authors provided final approval of the version to be published and agree to be accountable for all aspects of the work.

\bibliographystyle{unsrtnat}
\citestyle{acmnumeric}
\bibliography{references,nimo}

\appendix

\end{document}